\documentclass[sigconf]{acmart}

\usepackage{algorithm}
\usepackage{algorithmic}
\usepackage{graphicx}
\usepackage{textcomp}
\usepackage{xcolor}
\usepackage{stfloats}
\usepackage{subcaption}
\usepackage{booktabs}
\usepackage{pgfplots}
\usepackage{makecell}
\usepackage{hyperref}
\usepackage{multirow}
\usepackage{enumitem}
\usepackage{lipsum}
\usepackage{wrapfig}
\usepackage{tikz}
\usetikzlibrary{positioning,fit,decorations.pathreplacing}

\AtBeginDocument{%
  }
\acmConference[Anonymous Submission]{Conference}{2024}{Location}

\newif\ifcomment
 \commenttrue

 \newcommand{\rev}[1]{\textcolor{black}{#1}}

 \renewcommand\footnotetextcopyrightpermission[1]{} 
\begin{document}

\title{EXAM: Exploiting Exclusive System-Level Cache in Apple M-Series SoCs for Enhanced Cache Occupancy Attacks}
\thanks{This paper has been accepted to ACM ASIA CCS 2025 and will appear in the proceedings.}

\author{Tianhong Xu}
\affiliation{%
  \institution{Northeastern University}
  \city{Boston}
  \state{Massachusetts}
  \country{USA}}
\email{xu.tianh@northeastern.edu}

\author{Aidong Adam Ding}
\affiliation{%
  \institution{Northeastern University}
  \city{Boston}
  \state{Massachusetts}
  \country{USA}}
\email{a.ding@northeastern.edu}

\author{Yunsi Fei}
\affiliation{%
  \institution{Northeastern University}
  \city{Boston}
  \state{Massachusetts}
  \country{USA}}
\email{y.fei@northeastern.edu}

\begin{abstract}
Cache occupancy attacks exploit the shared nature of cache hierarchies to infer a victim's activities by monitoring overall cache usage, unlike access-driven cache attacks that focus on specific cache lines or sets.
There exists some prior work that target the last-level cache (LLC) of Intel processors, which is inclusive of higher-level caches, and L2 caches of ARM systems.
In this paper, we target the System-Level Cache (SLC) of Apple M-series SoCs, which is exclusive to higher-level CPU caches. We address the challenges of the exclusiveness and propose a suite of SLC-cache occupancy attacks, the first of its kind, where an adversary can monitor GPU and other CPU cluster activities from their own CPU cluster.
We first discover the structure of SLC in Apple M1 SOC and various policies pertaining to access and sharing through reverse engineering.
We propose two attacks against websites. One is a coarse-grained fingerprinting attack, recognizing which website is accessed based on their different GPU memory access patterns monitored through the SLC occupancy channel.
The other attack is a fine-grained pixel stealing attack, which precisely monitors the GPU memory usage for rendering different pixels, through the SLC occupancy channel.
Third, we introduce a novel screen capturing attack which works beyond webpages, with the monitoring granularity of 57 rows of pixels (there are 1600 rows for the screen).
 This significantly expands the attack surface, allowing the adversary to retrieve any screen display, posing a substantial new threat to system security.
  Our findings reveal critical vulnerabilities in Apple's M-series SoCs and emphasize the urgent need for effective countermeasures against cache occupancy attacks in heterogeneous computing environments.
  \end{abstract}


\maketitle

\section{Introduction}
Cache attacks~\cite{kayaalp2016high,liu2015last,f+r2014,percival2005cache,wang2006covert,yao2018coherence} exploit microarchitectural features of modern processors, such as timing differences between cache hits and misses, to leak sensitive information. 
Cache occupancy attacks~\cite{shusterman2020website,shusterman2021prime+,cronin2021exploration} are a nuanced variant that targets the overall state of cache occupancy to infer sensitive information, breaching privacy and confidentiality. The popular access-driven cache attacks focus on specific cache sets or lines to infer memory address-related secrete. 

ARM's big.LITTLE architecture exemplifies heterogeneous computing systems by combining high-performance ``big'' cores with energy-efficient ``LITTLE'' cores. ARM introduced the System Level Cache (SLC), an exclusive last-level cache situated between the higher-level caches (L1 and L2) and main memory~\cite{arm2022corelink,armDynamIQManual}. 
Apple's M-series System-on-Chips (SoCs) build upon ARM's heterogeneous designs, featuring multiple CPU clusters with local caches, an integrated GPU, and an SLC shared among clusters and the GPU. While this architecture offers significant performance benefits, it also introduces new security challenges due to the shared SLC between clusters and GPU.

Previous studies have explored microarchitectural side-channels in heterogeneous systems. Dutta et al.\cite{dutta2021leaky} and Almusaddar et al.\cite{almusaddar2023exploiting} demonstrated covert channels between CPUs and integrated GPUs in Intel systems by exploiting shared caches. Kou et al.\cite{kou2022attack} revealed side-channel attacks based on snoop filters in ARM processors. Cronin et al.\cite{cronin2021exploration} discussed the potential of cache occupancy attacks on ARM's SLC but did not delve into its structure or the unique challenges it presents.

In this paper, we present a novel suite of cache occupancy attacks targeting the SLC of Apple M-series SoCs—the first to exploit an \textit{exclusive} last-level cache, where an adversary can monitor GPU and other CPU cluster activities from their own CPU cluster. By reverse-engineering the SLC's sharing mechanism and structure in the Apple M1, we obtain critical insights that enable these attacks.
We demonstrate the effectiveness of the SLC occupancy side-channel through three attacks:

\begin{enumerate}
    \item \textbf{Website Fingerprinting Attack}: We perform a website fingerprinting attack, showing that our method achieves high precision across a wider range of scenarios, including cross-browser setups, where previous cache occupancy channels fail~\cite{shusterman2020website}.

    \item \textbf{Cross-Origin Pixel Stealing Attack}: We conduct a finer-grained cross-origin pixel stealing attack by exploiting the SLC occupancy channel. We leverage the data-dependent behavior of GPU rendering and compression, and accurately retrieve the screen display pixel-by-pixel, violating confidentiality and privacy.  Unlike previous attacks that relied on measuring the rendering time, our approach uses the SLC occupancy side-channel to extract pixel-level information. It is effective even in the presence of constant rendering-time implementations and Apple's recent security fixes addressing related vulnerabilities (CVE-2023-38599~\cite{CVE-2023-38599}).

    \item \textbf{Screen Display Snooping Attack}: Notably, we introduce a novel attack that does not rely on website-processing tools. By monitoring the GPU's rendering processes via the SLC occupancy channel, we can extract any information displayed on the screen. This significantly expands the attack surface beyond web pages, allowing the adversary to compromise any on-screen information, posing a substantial new threat to system security.
\end{enumerate}
\section{Background}
\subsection{Cache and its inclusion policies}
Caches are critical components in computer architecture for performance.
They are typically organized in hierarchical levels based on their proximity to the CPU, size, and speed to balance between access speed and capacity, with L1 being the smallest and fastest. 
Lower-level caches can be categorized into inclusive, exclusive, non-inclusive, and non-exclusive, each specifying how to store data across adjacent levels. 
For inclusive caches, the data stored in higher levels (e.g., L1) is also duplicated in lower levels (e.g., L2, L3), ensuring data evicted from one level remains accessible at other levels, simplifying the cache coherency. Exclusive caches prevent data duplication across levels, maximizing the cache space utilization. Other two policies offer middle-grounds with flexibility in cache management. 

\subsection{Cache side-channel attacks}
Cache side-channel attacks exploit the sharing nature of cache resources and the timing differences between cache hits and misses to retrieve secret information, such as cryptographic keys and private user information, without direct access to the victim's memory space, posing serious confidentiality threats in the user space.

\noindent\textbf{Access-driven attacks:} 
Traditional cache side-channel attacks, such as Prime+Probe \cite{liu2015last} and Flush+Reload \cite{f+r2014}, monitor the state of specific cache sets/lines to determine whether the victim has accessed them or not, to glean memory address information which is secret-dependent. 
These attacks often rely on high-resolution timers to measure minute differences in access times accurately. 

\noindent\textbf{Cache occupancy attack:} Cache occupancy attack, originally proposed by Oren et al. \cite{shusterman2020website}, differs significantly from access-driven cache attacks.  It focuses on monitoring the victim's contentions over the entire cache space, while prior access-driven attacks monitor contentions on selected cache sets/lines.  
Specifically, the spy allocates a buffer (equivalent to the size of the shared cache) and measures the time to access the entire buffer after victim execution.  If the victim's memory accesses cause some portion of the buffer to be evicted from the shared cache, the spy will recognize it with longer access time. The access time is roughly proportional to the number of cache lines that the victim uses. 

\subsection{Apple M-series SoCs and Their Cache Structures}

ARM's big.LITTLE technology represents a heterogeneous computing architecture designed to create more efficient processors. This architecture combines ``big'' core clusters and ``LITTLE'' core clusters, dynamically adjusting the processing elements based on computational demands to achieve both high performance and energy efficiency. The introduction of clusters in ARM's architecture has led to notable differences in cache structures from Intel. Specifically, in ARM architectures, the L1 cache is core-specific, the L2 cache is shared between cores of a cluster and is inclusive of the L1 cache, and a System Level Cache (SLC) is shared across clusters.

\begin{figure}[!t]
    \centering
    \includegraphics[width=0.5\textwidth]{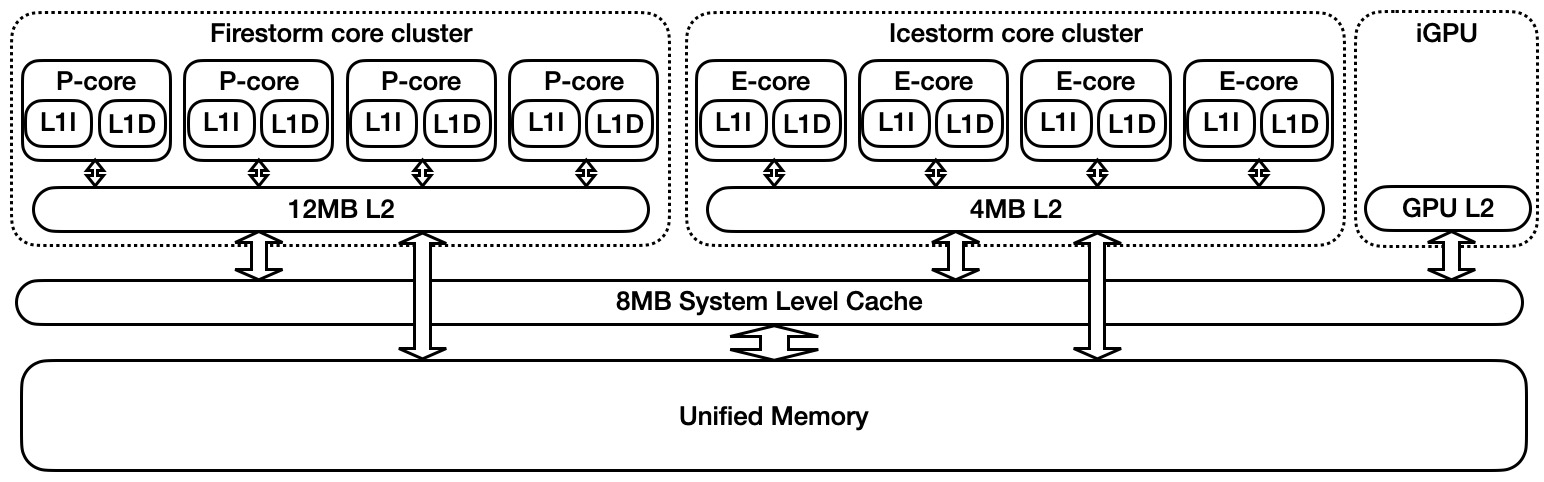}
    \caption{Cache structure of Apple M1}
    \label{m1cache}
    \vspace{-5mm}
\end{figure}

The Apple M-series SoCs, based on ARM's big.LITTLE architecture, herald a significant shift towards efficiency and performance in Mac computers. 
Figure~\ref{m1cache} shows the cache structure of the Apple M1. The M1 has four high-performance ``Firestorm'' cores (p-cores) and four energy-efficient ``Icestorm'' cores (e-cores), forming two CPU clusters. 
The SLC is shared across the clusters and GPU, and the SOC also includes a Unified Memory Architecture (UMA), streamlining access to a shared memory pool for both the CPU and GPU. 

Although Apple has not publicly disclosed the inclusive policy of the SLC, since in most M-series SoCs the size of the SLC is smaller than that of the CPU's L2 cache, it is unlikely that the SLC is  inclusive of the CPU caches. This creates significant differences from Intel's LLC, which is inclusive~\cite{backes2019impact}.  
Additionally, while we know that the SLC is shared between the CPU and GPU, its sharing mechanism has not been disclosed.

\section{SLC Occupancy Side-Channel on Apple Silicon}\label{sec3}

\subsection{Overview and challenges of the SLC occupancy channel}\label{sec31}
Current cache occupancy attacks~\cite{shusterman2020website} are implemented by accessing a memory buffer and measuring the total data access time, where the buffer is set at the size of a shared cache. For a cache hierarchy where all the lower levels are inclusive,
the common assumption is that the whole LLC can be filled by the buffer data and therefore the spy is able to measure the cache occupancy (evictions) by the victim.
However, this assumption does not hold for the SLC on Apple Silicon, as it is not inclusive to CPU's local caches.
The SLC cannot be directly filled by loading the buffer data.
When accessing a buffer element, the data, which is newer and warmer, is always loaded to the CPU's local caches (L1 and L2), not necessarily to the SLC.
Details on Apple M1 SLC's inclusive policy, sharing mechanism, and other sharing related policies are not publicly disclosed. This lack of information presents significant challenges in designing effective cache occupancy attacks targeting the SLC.

The prior work~\cite{shusterman2021prime+, cronin2021exploration} are the only two cache occupancy attacks on ARM processors. Shusterman et al.\cite{shusterman2021prime+} targets L2 cache of an Apple M1, the highest inclusive cache level, by setting the buffer at the size of L2 cache. This limits the spy to be on the same CPU cluster as the victim.
Although Cronin et al.\cite{cronin2021exploration} discussed the potential role of the ARM SLC in cache occupancy attacks, their attack design did not consider SLC's structure characteristics and its inclusive policy, and they did not specify whether they are measuring the cache occupancy of the SLC or L2.  Both the prior two attacks do not apply to inter-cluster and CPU-GPU scenarios.

In this section, we first quantify the access (hit) time for the Apple M1 SLC, allowing us to precisely differentiate SLC access from main memory access. We then revisit previous cache occupancy attacks and show that these attacks cannot monitor the contention state of the Apple M1 SLC, because their buffer access patterns do not cater to the SLC's inclusive policy. To address this issue, we propose a different access pattern. Next, we reverse engineer the SLC's inclusive policy, set indexing and replacement policy. Through our analysis, we discover that the SLC operates as exclusive with respect to the CPU caches but inclusive with respect to the GPU cache. Building upon these insights, we propose a novel SLC occupancy side-channel that efficiently monitors the contention state of the SLC. We compare the performance of our approach with previous cache occupancy channels, demonstrating its superiority in inter-cluster and cross CPU-GPU scenarios.

\subsection{Quantifying SLC Hit Latency}\label{subsec:SLC_hits}

\begin{wrapfigure}{r}{0.6\linewidth}
    \vspace{-3mm}
    \centering
    \includegraphics[width=0.95\linewidth]{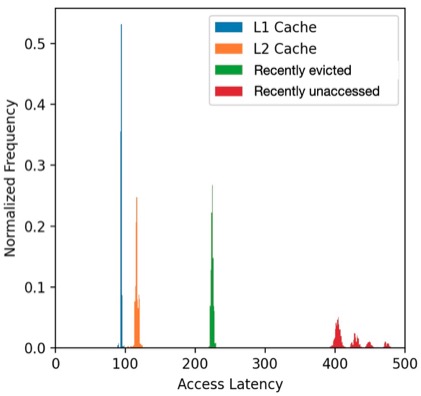}
    \caption{Hit time of different cache levels}
    \label{fig32}
    \vspace{-3mm}
\end{wrapfigure}
To establish a reliable SLC occupancy side-channel, we first need to recognize whether a piece of data is present in the SLC (i.e., an SLC cache hit) by measuring data access latencies. Prior works~\cite{hetterich2022branch,ravichandran2022pacman} have quantified L1 hit, L2 hit, and L2 miss times for the Apple M1. However, they have not differentiated between SLC hits and SLC misses, both of which are categorized as L2 misses.

Based on ARM's documentation~\cite{arm2022corelink,armDynamIQManual}, we speculate that the SLC acts as a backward storage for the CPU's L2 cache; that is, when a cache line is evicted from the CPU's L2 caches, it is spilled over to the SLC. To test this hypothesis, we design an experiment using an L2 eviction set~\cite{vila2019theory}. All experiments in this section are conducted on a MacBook Air equipped with an Apple M1 chip, running the macOS operating system. We employ a user-level counting-thread timer based on prior work~\cite{hetterich2022branch,ravichandran2022pacman}, which achieves a resolution of 3~GHz.

In our experiment, we first access a piece of data to ensure it is loaded into the cache hierarchy. We then employ an L2 eviction set to evict that data from the L2 cache. Immediately after it, we access the same data again and measure the access time. We observe that the access time of the data just evicted from the L2 cache is significantly shorter than that of data not recently loaded into the cache hierarchy (by using a sufficiently large eviction buffer to ensure this). Specifically, the access time for the newly evicted data is approximately 220 ticks, whereas the access time for much older data is around 430 ticks, as shown in Figure~\ref{fig32}.  We infer that the 220 ticks correspond to an SLC hit time. Figure~\ref{fig32} also presents the L1 cache hit time and L2 cache hit time. Correspondingly, we set two thresholds, 160 ticks and 300 ticks, to differentiate between local cache hits (L1 or L2), SLC hits, and SLC misses.

Additionally, we find that when data is first accessed by the GPU and then accessed by the CPU, the CPU's access time also aligns with the SLC hit time. This suggests that the SLC is shared between the CPU and GPU. 

    \subsection{Previous cache occupancy attacks}\label{L2}
   We next review the cache occupancy channels in prior work~\cite{shusterman2021prime+, cronin2021exploration, shusterman2020website}. We analyze how the access patterns allocate data from the buffer to different levels of caches on Apple M1. 
    Our findings reveal that the previous access patterns fail to consistently fill the SLC and cannot accurately measure the contention state within the SLC. To address this issue, we propose a simple yet effective improvement to the access pattern.

    \noindent\textbf{Access pattern of previous cache occupancy channels:}
    In previous ARM L2 cache occupancy attacks, the data structure of the buffer consists of a range of contiguous virtual addresses, with the number of data entries equal to the number of cache lines in the L2 cache, where the size of data entry is the cache line size (128 bytes).
    The access pattern, which determines the order in which these data entries are accessed, have been intricately designed to circumvent the effects of hardware prefetchers, which can otherwise mask true cache behavior by pre-loading anticipated data.
    However, these approaches did not take into consideration of the L2 cache set indexing. Prior work~\cite{yu2023synchronization} shows that the L2 cache set indexing of Apple M1 is tailored for optimal performance, where 11 bits of the cache set index are mapped directly from the large-page memory address offset and two upper bits are the XOR-ed result of the page number.
    Consequently, the previous cache occupancy attacks could not evenly fill all the L2 cache sets with their chosen buffer size and access pattern. This uneven distribution often results in some cache sets becoming overflown—exceeding their capacity of 12 ways—while others remaining underutilized.

    \noindent\textbf{Problems in profiling phase:}
    For cache occupancy attacks, the adversary keeps accessing the buffer with a certain access pattern and measuring the total access time, where we define each run a ``profiling'' process, and both the traditional cache ``priming'' and ``probing'' functions are embodied in profiling.
    The purpose of priming is to precisely fill the L2 cache with the buffer data, evenly populating each set.  However, due to the aforementioned uneven distribution, approximately half of the L2 cache sets remain unfilled while the other half are overflown.
    For an unfilled L2 set, the victim's memory access will only cause contentions on the spy's cache occupancy when the victim data exceeds the remaining capacity of the set, i.e., resulting in undercount of the victim data access.
    For an overflown set, the problem is more severe with a phenomenon we define as self-eviction. During the previous profiling, the buffer data that fills an overflown L2 set is the last (newest) to be accessed, while the data accessed first has been evicted to the SLC. Since each profiling follows the same access order, the early data access results in a SLC hit with the data being copied to the L2 cache, which will evict some L2 cache lines,  resulting in L2 misses for subsequent accesses. Consequently, regardless of whether the victim has memory access or not, the spy data self-evictions result in L2 cache misses, i.e., overcounting if attributing L2 cache misses to victim data accesses.

If we increase the buffer size, approaching the SLC size and  measure the SLC hit times, the self-eviction may also cause problems for sequential order. At the end of a profiling phase, some old data is in SLC and new data is in the L2 cache. For the next profiling phase, as the access order is the same, the early accessed data is brought to L2 cache and evict some data to SLC, which when accessed will be counted as SLC hits, even though they are not really in the SLC before this profiling phase. This will cause overcounting of SLC hits.

\begin{figure}[t]
        \centering
        \begin{subfigure}[b]{0.5\columnwidth}
            \includegraphics[width=\linewidth]{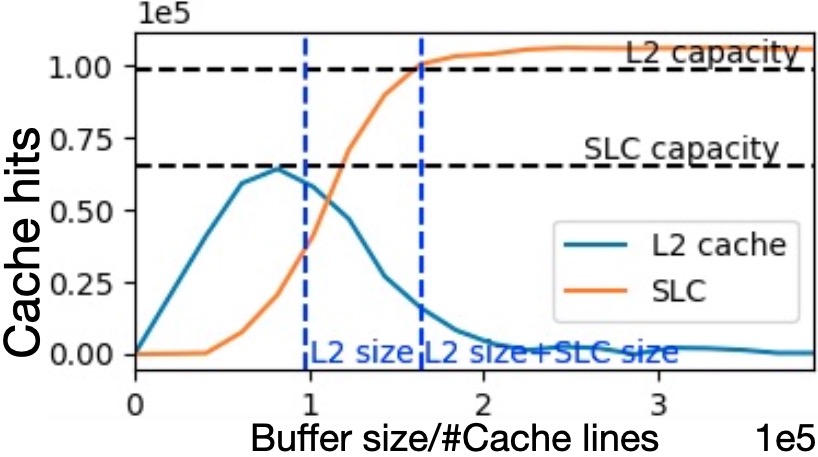}
            \caption{Sequential-order}
            \label{fig32:a}
        \end{subfigure}%
        \hfill
        \begin{subfigure}[b]{0.5\columnwidth}
            \includegraphics[width=\linewidth]{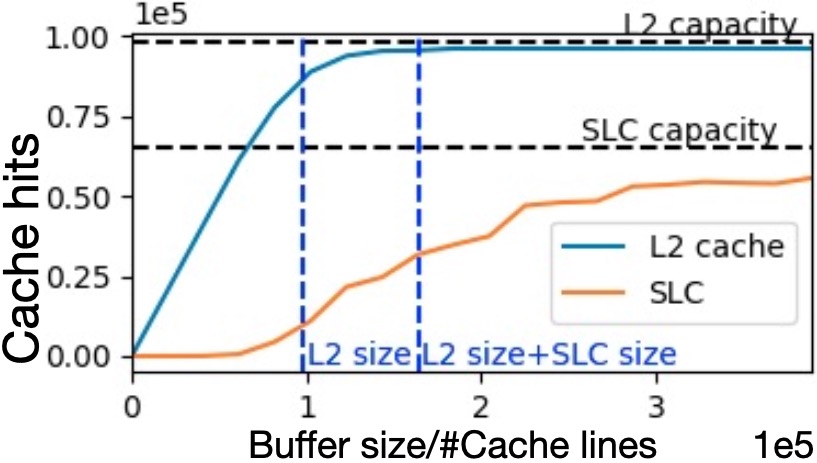}
            \caption{Alternated-order}
            \label{fig32:b}
        \end{subfigure}
        \caption{L2 \& SLC hits under different access patterns}
        \label{accesspattern}%
        \vspace{-5mm}
    \end{figure}

    Figure~\ref{fig32:a} illustrates the relationship between the measured number of L2 cache hits and SLC cache hits and the buffer size under this sequential access pattern. When the buffer is small, all data fits within the L2 cache without overflowing any cache set, resulting in a linear relationship between the buffer size and the number of L2 cache hits. However, as the buffer size approaches half of the L2 cache capacity, some cache sets begin to overflow, leading to a sharp decline in the number of L2 hits and an increase in SLC hits. Due to self-eviction, we can observe that as the buffer size continues to grow, L2 hits gradually decrease to nearly zero, indicating a significant undercount. On the other hand, SLC hits increase to a value far exceeding the SLC's capacity, suggesting a substantial overcount. These miss-counting issues persist at any buffer size, making it challenging to accurately assess the contention status of L2 or SLC based on their hit counts.

    \noindent\textbf{Alternated-order access pattern:}
    To address these issues, we propose a simple adjustment in the access pattern: the access order alternates between sequential and reverse between two consecutive profiling phases.
    This ensures that for an overflown cache set, the data that was last inserted during the previous profiling cycle is accessed first in the current cycle. According to the LRU policy, this previously last-entered data remains in the local cache and accessing it results in cache hits. Therefore, the attacker can accurately detect and monitor the state of the L2 cache.

    Figure~\ref{fig32:b} illustrates the relationship between the number of L2 and SLC hits and the buffer size under the alternated-order access pattern. As the buffer size increases, the number of L2 hits and SLC hits gradually approach their respective cache capacities.
    This behavior indicates that both the L2 cache and SLC can effectively reflect their respective contention statuses under this access pattern.

    The alternated-order access pattern proves crucial for our subsequent analyses and attacks. By mitigating self-eviction effects and providing more accurate measurements of cache occupancy, this approach enables us to precisely quantify the total capacity and contention within the SLC. 

    \subsection{Understanding the structure and policies of M1  SLC}\label{reverse}

    \rev{Given the limited public information about the SLC on Apple's M-series SoCs, we reverse-engineered its underlying structure and policies. Due to page limitations, this section focuses on presenting the conclusions, while the detailed experimental process is provided in Appendix \ref{AppendixA}.}

    \noindent\textbf{Cache line size: }Our first observation is that system-level cache line size is 128 bytes, the same as L2 caches.

    \noindent\textbf{Inclusiveness Policy:}
Our experiments reveal that the SLC in Apple M1 employs a hybrid inclusiveness policy: inclusive with respect to the GPU cache but exclusive with respect to the CPU cache. This unique configuration optimizes performance while maintaining coherence across the heterogeneous system. \rev{We derived this conclusion from carefully designed experiments, detailed in Appendix \ref{InclusivenessPolicy}.}

    \noindent\textbf{SLC set index mapping:}
    We discover a distinctive SLC set index mapping mechanism. Unlike typical cache configurations that utilize lower bits of the memory address for cache indexing, M1’s SLC excludes the lowest 13 bits of the physical address for indexing and uses bits from the 14th position and above. \rev{The detailed experiments and results are shown in Appendix \ref{SetindexPolicy}.}

    \noindent\textbf{Replacement policy:} 
    Our observation suggests that the SLC's replacement policy is independent of access order, indicating a pseudo-random policy.  \rev{The detailed experiments and results are shown in Appendix \ref{ReplacementPolicy}.}

    \subsection{Occupancy side-channels on Apple M1}\label{SLC}
    Building on our reverse engineering insights into the SLC's structure and behavior, we now propose a novel SLC occupancy channel. This channel is specifically designed to exploit the unique characteristics of the M1 SLC uncovered, particularly its exclusive nature with respect to CPU caches. Our key contribution is the development of an SLC occupancy side-channel that strategically bypasses the L2 cache, directly targeting the SLC and addressing the challenges posed by its architecture.

    To comprehensively evaluate the effectiveness of our new SLC occupancy channel, we compare its performance against two more conventional occupancy channels: the L2 occupancy channel and the total occupancy channel. Our comparison spans various scenarios, including intra-cluster, inter-cluster, and CPU-GPU interactions, providing a thorough assessment of each channel's capabilities and limitations, with a particular focus on demonstrating the advantages of our SLC-specific one.

    In this section, we present and analyze three distinct occupancy channel implementations:

    \noindent\textbf{L2 Cache Occupancy:} In this side-channel, the spy fills the entire p-cores' L2 cache with its buffer and measures the L2 cache occupancy. We use the same data structure as the prior work \cite{shusterman2020website} and employ an alternated-order access pattern.The results of Figure \ref{fig32:b} show that the L2 cache size is 120,000 cache lines. 
    This side-channel can only monitor the contention state of the L2 cache, and cannot monitor the SLC status. 

    \noindent\textbf{Total Occupancy:} This side-channel aims to completely fill both the L2 cache (p-cores) and the SLC, to the greatest extent possible. We still use the same data structure and access pattern as in the L2 cache occupancy channel, and
    set the buffer size at 300,000 cache lines, which is the point where the SLC reaches its utilization limit, as shown in Figure \ref{fig32:b}. Theoretically, this-side channel should be able to monitor the contention status of both the L2 cache and SLC. However, due to the large buffer size, the time for one profiling in this channel is very long, resulting in a very limited sample rate and large noise.

\begin{figure}[t]
        \centering
        \begin{subfigure}[b]{0.5\columnwidth}
            \includegraphics[width=\linewidth]{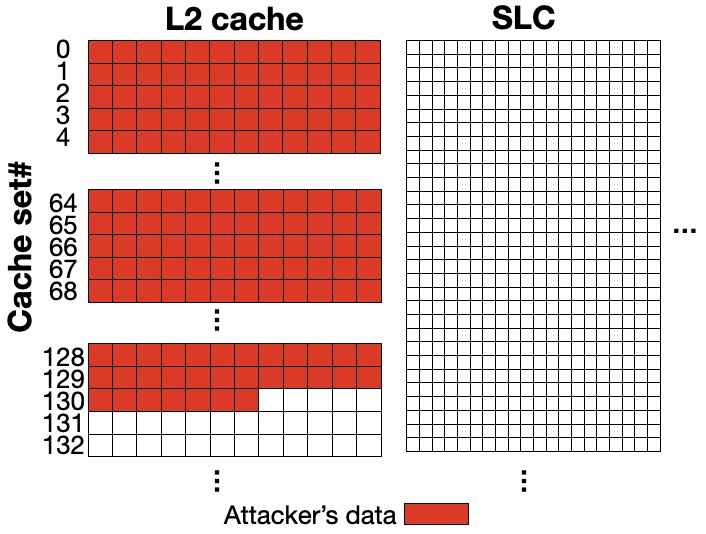}
            \caption{stride=128}
            \label{fignew3a}
        \end{subfigure}%
        \hfill
        \begin{subfigure}[b]{0.5\columnwidth}
            \includegraphics[width=\linewidth]{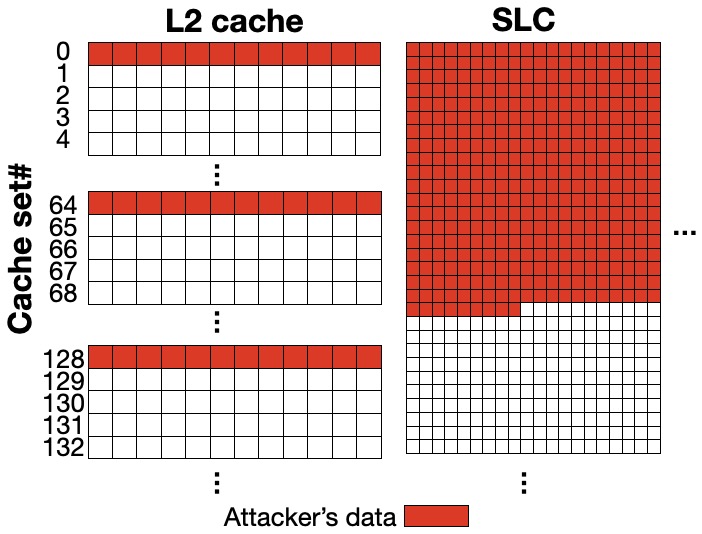}
            \caption{stride=8192}
            \label{fignew3b}
        \end{subfigure}
        \caption{Cache filling with different data structure}
        \label{fignew3}%
        \vspace{-5mm}
    \end{figure}

    \noindent\textbf{SLC Occupancy:} This side-channel aims to fill only the SLC with a buffer. We propose a new data structure that strategically circumvents the L2 cache to directly fill the SLC. This approach leverages the difference in set-indexing between the L2 cache and the SLC: while the L2 cache uses the lowest 13 address bits for indexing, the SLC does not, allowing us to manipulate the address stride to effectively restrict accesses on the L2 cache to certain sets.
     The new data structure employs an a range of contiguous virtual addresses with a stride of 8192, which make their lowest 13 address bits fixed, enabling us to fill the SLC using only a small portion of the L2 cache, i.e., more efficient profiling.
     \rev{Figure~\ref{fignew3} shows the cache utilization patterns of both traditional (stride=128) and our new data structure (stride=8192) approaches. With traditional data structure, data fills L2 evenly, making it difficult to isolate SLC behavior. In contrast, our new approach restricts the L2 cache usage to just 1/64 of its capacity, approximately 192 KB, while still engaging majority of SLC cache lines. We set the buffer size to 80,000 cache lines, which is the threshold value just before the number of L2 cache hits reaches their peak, as shown in Figure~\ref{fig32:b}. }

The contention status of the L2 cache is affected by victim processes located on the same cluster (p-cores) as the spy: any victim memory accesses will evict some of the spy's data out of the L2 cache. The contention status of the SLC is not as intuitive because of its exclusiveness as explained before.
    Theoretically, the L2 cache occupancy channel only works for intra-cluster scenarios,  while the other two channels can monitor inter-cluster or CPU-GPU scenarios.
      In our experimental setup,
   we let the attacker do the profiling every 50 milliseconds, and introduce victim activities that involve loading a victim buffer in the interval between two profiling phases. The size of the victim buffer is varied systematically across different trials to examine how these variations influence the attacker's profiling times. Each configuration for the victim buffer size is tested 100 times to gather a robust dataset, from which we calculate the variance and average values of the profiling times.
We pin the spy and victim processes specifically to different cores to create the following three conditions:

\noindent\textbf{Intra-cluster}: Both the victim and attacker processes are run on the high-performance ``Firestorm'' cores.

\noindent\textbf{Inter-cluster}: The victim and attacker operate on different CPU clusters, with the victim on the energy-efficient ``Icestorm'' cores and the attacker on the ``Firestorm'' cores.

\noindent\textbf{CPU-GPU}: The attacker runs on a CPU core ("Firestorm"), while the victim process is exclusively a GPU process.

\begin{figure}[htb]
    \centering
    \begin{subfigure}[b]{0.15\textwidth}
        \centering
        \includegraphics[height=1.6\textwidth]{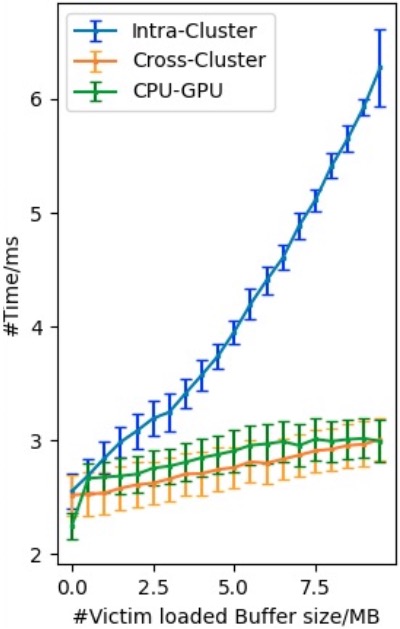}
        \caption{L2 Occupancy}
        \label{fig:sub1}
    \end{subfigure}
    \hfill
    \begin{subfigure}[b]{0.15\textwidth}
        \centering
        \includegraphics[height=1.6\textwidth]{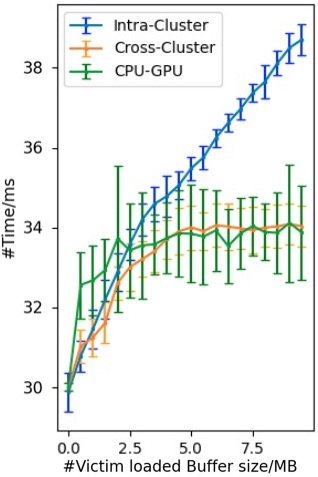}
        \caption{Total Occupancy}
        \label{fig:sub2}
    \end{subfigure}
    \hfill
    \begin{subfigure}[b]{0.15\textwidth}
        \centering
        \includegraphics[height=1.6\textwidth]{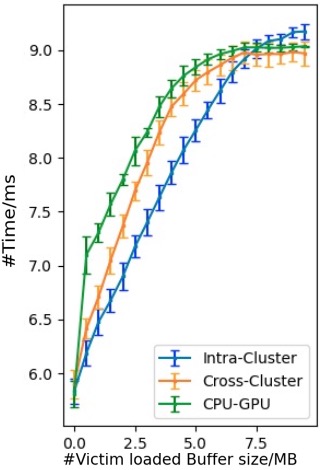}
        \caption{SLC Occupancy}
        \label{fig35:c}
    \end{subfigure}
    \caption{Spy access time vs. victim activities}
    \label{fig35}%
\vspace*{-0.4cm}
\end{figure}

Figure~\ref{fig35} presents our experimental results across different scenarios, comparing the three occupancy channels. The x-axis represents the victim's buffer size, while the y-axis shows the spy's profiling time. An effective channel should exhibit a linear relationship between these variables, with the slope indicating the side-channel signal strength. Our results demonstrate that the SLC occupancy channel outperforms both the L2 and Total occupancy channels, providing reliable and efficient monitoring across all tested configurations, including inter-cluster and CPU-GPU scenarios where the L2 channel fails.

Our eviction set implementation is designed to address two critical challenges specific to SLC occupancy attacks: efficiently filling the SLC and bypassing the L2 cache. This approach optimizes for overall cache usage monitoring, which is the primary focus of occupancy attacks, rather than extracting fine-grained spatial information. 
This design choice is particularly advantageous for the threat models explored in subsequent chapters, where attackers may have limited capabilities to construct precise eviction sets. In the following sections, we showcase the power and versatility of our SLC occupancy channel through three distinct attacks spanning website fingerprinting, cross-origin website pixel stealing, and screen capturing. These practical demonstrations not only validate our approach but also highlight the broad applicability of this attack vector, emphasizing the urgent need for security measures in modern heterogeneous computing environments. 
\section{Website fingerprint attack}\label{sec:sec4}
Our SLC occupancy side-channel shows effectiveness in various scenarios, particularly for inter-cluster and cross CPU-GPU situations, opening up more attack surfaces.
We apply such SLC cache occupancy side-channels against real applications, and build a new website fingerprinting attack.

\noindent\textbf{Threat model:}
\rev
{Following prior work \cite{oren2015spy,cronin2021exploration,shusterman2021prime+,shusterman2018robust}, this attack employs malicious JavaScript -- delivered via ads or websites -- running unobtrusively in the background to monitor a victim user's web browsing activity. Constrained by standard JavaScript APIs and low-resolution timers, the attacker operates without exploiting browser vulnerabilities. Assuming  browsing traces under identical conditions (browser and device) are collected before-hand, the attacker profiles them, then monitors the victim's real-time behavior, and correlate the target website's trace with the profiling traces. Furthermore, we extend the prior attack framework to encompass cross-browser scenarios, where the attacker's code executes in one browser and the victim uses a different browser.
}

\subsection{Experimental Setup}
Website fingerprinting attack aims to covertly track a user's website browsing history to facilitate malicious activities such as targeted advertising or blackmailing. We build website fingerprinting attacks on MacOS systems across three popular browsers: Chrome, Safari, and Firefox.  We evaluate both cross-tab (same browser) and cross-browse scenarios, where the spy (tab or browser) runs in the background and the victim browses in the foreground.
To evaluate the generalizability of our findings, we test three different Apple devices,  with M1, M1 Pro, and M3 Pro SoCs, respectively, which differ in the core, cluster, and SLC configurations.
Across all experimental configurations, we evaluate the performance of two side-channel approaches: the previous L2 cache occupancy channel~\cite{shusterman2021prime+} and our SLC occupancy channel.

\subsection{SLC occupancy attack validation}\label{vali}
We first explore the capabilities of the SLC side-channel in monitoring website activities compared to the previous cache occupancy attacks. 
We design a benchmark with typical, but controllable, website behaviors, which allows us to systematically analyze the side-channel's sensitivity and accuracy.

\noindent\textbf{Rendering benchmark}: We design the benchmark to simulate the visual rendering activities of a website, which include loading and updating of text, images, and the overall layout. 
We employ image loading as the primary task that involves dynamically fetching and displaying images.
In this benchmark, we make random decisions at fixed intervals to determine whether to perform the image loading operation. We record such decisions as binary values, with 1 indicating loading image and 0 meaning no-loading. These intervals are synchronized with the spy's sampling rate. Specifically, we run a profiling phase for every 50ms for a total duration of 20 seconds,resulting in a side-channel trace of 400 measurements. 
By calculating the correlation between the side-channel trace and the benchmark behavior trace, we can assess the strength of the side-channel.

We employ the T-test to evaluate the correlation~\cite{kim2015t, boneau1960effects}.
T-test is a statistical method used to compare the means of two sets of data and determine if there is a significant difference between them. In our experiment, we divide all the data points in the side-channel trace into two groups based on the corresponding values (0 or 1) in the benchmark behavior trace. 
A higher T-test score indicates a greater difference between the two groups, implying a stronger correlation between the side-channel trace and the victim activity. Typically, a T-test score above 4.5 is considered statistically significant and used as the threshold.

\begin{table}[!htbp]
  \centering
  \label{tab:cache-measurements}
  \begin{tabular}{l|cc|cc|cc}
  \hline
  \textbf{Browsers} & \multicolumn{2}{c|}{\textbf{M1}} & \multicolumn{2}{c|}{\textbf{M1 Pro}} & \multicolumn{2}{c}{\textbf{M3 Pro}} \\
  Spy-Victim & \textbf{SLC} & \textbf{L2~\cite{shusterman2021prime+}} & \textbf{SLC} & \textbf{L2} & \textbf{SLC} & \textbf{L2} \\
  \hline
  Cross-tab Chrome     & 26.1 & 27.2 & 23.2 & 22.1 & 24.2 & 5.2 \\
  Cross-tab Safari     & 11.6 & 7.3  & 11.2 & 6.6  & 12.5 & \textcolor{red}{2.4} \\
  Cross-tab FireFox    & 13.8 & 15.7 & 14.7 & 6.8  & 14.5 & 6.7 \\
  Chrome-Safari        & 14.3 & 11.6 & 12.3 & \textcolor{red}{2.8}  & 14.2 & 5.8 \\
  Chrome-FireFox       & 15.2 & 13.5 & 14.8 & \textcolor{red}{4.4}  & 14.8 & 4.8 \\
  Safari-Chrome        & 14.6 & \textcolor{red}{2.8}  & 15.9 & \textcolor{red}{2.2}  & 15.2 & \textcolor{red}{2.8} \\
  Safari-FireFox       & 9.3  & \textcolor{red}{3.1}  & 7.5  & \textcolor{red}{3.6}  & 9.4  & \textcolor{red}{3.1} \\
  FireFox-Chrome       & 8.4  & 16.2 & 7.3  & \textcolor{red}{4.8}  & 9.8  & \textcolor{red}{4.4} \\
  FireFox-Safari       & 10.4 & 12.3 & 12.1 & \textcolor{red}{3.7}  & 10.1 & \textcolor{red}{2.4} \\
  \hline
  \end{tabular}
  \caption{Benchmark results/T-test score}
  \label{Rendering}
\end{table}

\vspace*{-5mm}

The experimental results are shown in Table~\ref{Rendering}. 
Overall, the SLC occupancy channel demonstrates correlations with benchmark activities  (T-values above 4.5 in all scenarios), while the previous L2 cache occupancy channel, although exhibiting similar correlations in some cases, proves to be uncorrelated with benchmark activities in others. We highlight those cases in red. As MacOS does not allow users to assign browser or tab processes to a specific core or cluster, we can only infer the allocation of background and foreground webpages to the processing clusters based on the experimental results. On an Apple M1, the prior L2 side-channel~\cite{shusterman2021prime+} becomes ineffective when the attacker operates from Safari in the background while the victim uses other browsers in the foreground.
As Safari is Apple's own browser, it may be uniquely optimized for Apple's M1 heterogeneous core architecture and  is relegated to an energy-efficient core when it runs in the background. Other browsers persist in high-performance cores even when they are in the background. This finding aligns with the conclusions drawn by Cronin et al.\cite{cronin2021exploration}. Only in this situation the attacker and the victim are placed on different clusters, forming an inter-cluster scenario. On an M1 Pro, the prior L2 side-channel does not work in all cross-browser scenarios. We speculate that it is due to M1 Pro having two clusters composed of performance cores, causing any two simultaneously running browsers to be placed on different clusters. On an M3 Pro, the prior L2 side-channel is even ineffective for cross-tab on Safari, and cross-browser when the spy is Firefox. 
We hypothesize that this may be the result of M3 Pro employing a more advanced process-core allocation strategy, leading to the attacker and victim being assigned to different clusters in these specific scenarios.

\subsection{Website fingerprinting attack}\label{sec43}
We next evaluate website fingerprinting attacks on a collection of realistic websites, based on the two cache occupancy side-channels.
We select Chrome-Chrome and Safari-Chrome as representatives of intra-cluster and inter-cluster scenarios, respectively, based on the similar results observed across the three devices in Section~\ref{vali}. 

\noindent\textbf{Data Sets:}
For our experiments, we utilize a closed-world dataset. We collect a dataset of 10,000 traces from the Alexa Top 100 websites \rev{for each combination of device and scenario}, with 100 traces per website and each trace lasting 8 seconds.
\rev{All the data is collected in a clean environment where no applications other than the attacker's and victim's browsers are running, and no other browser tabs are open in the background.}
For our SLC occupancy channel, we set the sampling rate to 10ms, while for the previous L2 side-channel, we adopt the same 2ms sampling rate~\cite{shusterman2021prime+}. The longer sampling rate for the SLC side-channel is due to its longer profiling time compared to the previous L2 channel, as shown in Figure~\ref{fig35}. \rev{The complete list of websites is given in Appendix B}.

\noindent\textbf{Machine Learning Approaches:} We employ Support Vector Machines (SVM) \rev{with a linear kernel} as the classifier on the collected side-channel traces. \rev{For feature extraction, we use the raw time-series from each trace - 800 sampling points for the SLC occupancy channel and 4000 points for the L2 occupancy channel. Despite the SLC occupancy channel having fewer sampling points, our experiments demonstrate that this does not affect the classification accuracy much. The datasets are normalized using Min-Max scaling to ensure consistent feature ranges.} The classifiers are trained and tested using a 90-10 cross-validation strategy. We report the Top-1 accuracy as the performance metric.
\begin{table}[ht]
    \centering
     \caption{Website fingerprinting accuracies of different side-channels on various SoCs under different scenarios}
    \label{side_channel_accuracy}
    \begin{tabular}{lllc}
    \toprule
    SoC         & Scenario       & Side-channel   & Accuracy \\
    \midrule
    \multirow{4}{*}{Apple M1}    & \multirow{2}{*}{Chrome-Chrome} & SLC  & 90.5\%     \\\cline{3-4}
     &  & L2~\cite{shusterman2021prime+}   & 91.2\%     \\\cline{2-4}
        & \multirow{2}{*}{Safari-Chrome} & SLC  & 87.4\%         \\\cline{3-4}
      &  & L2   & \textcolor{red}{33.4\%}     \\\hline
    \multirow{4}{*}{Apple M1 Pro} & \multirow{2}{*}{Chrome-Chrome} & SLC & 92.3\%     \\\cline{3-4}
     &  & L2  & 91.7\%     \\\cline{2-4}
     & \multirow{2}{*}{Safari-Chrome} & SLC & 88.6\%     \\\cline{3-4}
     &  & L2  & \textcolor{red}{35.3\%}     \\\hline
    \multirow{4}{*}{Apple M3 Pro} & \multirow{2}{*}{Chrome-Chrome} & SLC & 92.4\%     \\\cline{3-4}
   &  & L2  & 76.3\%     \\\cline{2-4}
      &  \multirow{2}{*}{Safari-Chrome} & SLC & 90.4\%         \\\cline{3-4}
     &  & L2  & \textcolor{red}{37.9\%}         \\
    \bottomrule
    \end{tabular}
       \end{table}

    Table~\ref{side_channel_accuracy} presents the experimental results, which align with our expectations. \rev{In the Chrome-Chrome scenario (cross-tab) on the same cluster, the SLC occupancy channel yields high accuracies of over $90\%$, which are similar to the previous L2 side-channel~\cite{shusterman2021prime+} on Apple M1 and Apple M1 Pro devices but the L2 side-channel accuracy is lower at $76\%$ on Apple M3 Pro. Furthermore, in the Safari-Chrome scenario, which represents a cross-cluster setting, the SLC occupancy channel maintains high accuracies of close to $90\%$ while the accuracy of the previous L2 side-channel significantly decreases.} 
    These findings demonstrate the effectiveness of the SLC occupancy channel in a wider range of application scenarios, highlighting its ability to capture the contention state of the shared SLC, regardless of the specific cluster allocation of the attacker and victim processes.

    \rev{We further evaluate our attack in real-world scenarios beyond the clean environment. We observe that most background applications (e.g., Preview, VS Code) and browser tabs displaying static content (e.g., Wikipedia pages) have negligible impact on the side-channel trace collection. The models trained in clean environments maintain similar accuracy when deployed in these scenarios. However, memory-intensive background applications and webpages, such as video conferencing tools (e.g., Zoom, Teams) and video playback applications, saturate the profiling phase timing to its maximum value, making collection of meaningful side-channel traces hard. }

\section{Cross-Origin Pixel Stealing Attack via SLC Occupancy Side-Channel}\label{sec5}

In this section, we demonstrate a novel cross-origin pixel stealing attack that leverages the SLC occupancy side-channel on Apple M-series SoCs. Our attack exploits the data-dependent nature of GPU compression~\cite{wang2024gpu}, allowing us to infer individual pixel values from a cross-origin iframe, even in the presence of constant-time SVG filter implementations and recent security mitigations (CVE-2023-38599) \cite{CVE-2023-38599}.

\subsection{Attack Overview}
Pixel stealing attacks exploit side-channels in web browsers to infer the values of individual pixels from cross-origin content, bypassing the same-origin policy. These attacks target SVG filters—the graphical operations applied to web graph content—to monitor and analyze iframe rendering times through the SLC occupancy channel, distinguishing between black and white pixels.

Previous approaches relied on measuring the SVG filter rendering times to differentiate between black and white pixels. GPU-zip\cite{wang2024gpu} introduced a technique using the LLC walk time on Intel systems to distinguish pixel colors, in addition to the traditional rendering time measuring method.
However, this approach cannot apply to Apple M1 systems due to differences between ARM's SLC and Intel's LLC Cache (LLC), as discussed in Section ~\ref{sec31}. For Apple M1 systems, GPU-zip can only rely on the SVG filter rendering time.

Due to the data-dependent nature of GPU compression, the SVG filter processes white and black iframes differently. Specifically, when handling a white iframe, the GPU memory usage is higher than when processing a black iframe.
By monitoring the GPU memory usage through the SLC occupancy side-channel introduced in Section \ref{SLC}, we can infer whether the target pixel is black or white based on the observed memory usage.

\noindent\textbf{Threat model:} 
\rev{
We assume a controlled environment in which an attacker-controlled webpage embeds a victim webpage as a cross-origin iframe. The attacker aims to extract sensitive visual information from the victim page displayed within the iframe, by inferring individual pixel values. This attack scenario requires the victim to visit the attacker's malicious webpage while she is authenticated to access sensitive target websites, with content such as usernames and profile pictures accessible via cookies or session data. The attack relies on the victim's browser supporting SVG filters and allowing cross-origin iframes.}

\noindent\textbf{Attack Setup:} 
\rev{
Similar to previous works~\cite{taneja2023hot, wang2024gpu}, our pixel-stealing attack employs carefully designed SVG filters to process individual pixels from a cross-origin iframe. The attack works by embedding the target cross-origin webpage within an iframe on the attacker's page and magnifying a specific target pixel to occupy a significant portion of the webpage. We apply SVG filters crafted to create pattern-dependent workloads that respond differently to black and white pixels. By observing side-channel effects through the SLC occupancy channel, we can infer the color of the original pixel. Similar to Section~\ref{sec43}, all experiments in this section were conducted in a clean environment, with no other applications running in the background, to ensure precise measurements.}

\subsection{Validating SLC Occupancy Channel for Pixel Color Discrimination}

To demonstrate the efficacy of the SLC occupancy channel in distinguishing black and white pixels, we conducted a series of experiments and compare this channel against the L2 cache occupancy channel. The experimental setups are same as Section~\ref{SLC} and Section~\ref{sec:sec4}.

\begin{figure}[htb]
    \centering
    \begin{minipage}{.48\linewidth}
        \centering
        \includegraphics[width=0.9\textwidth]{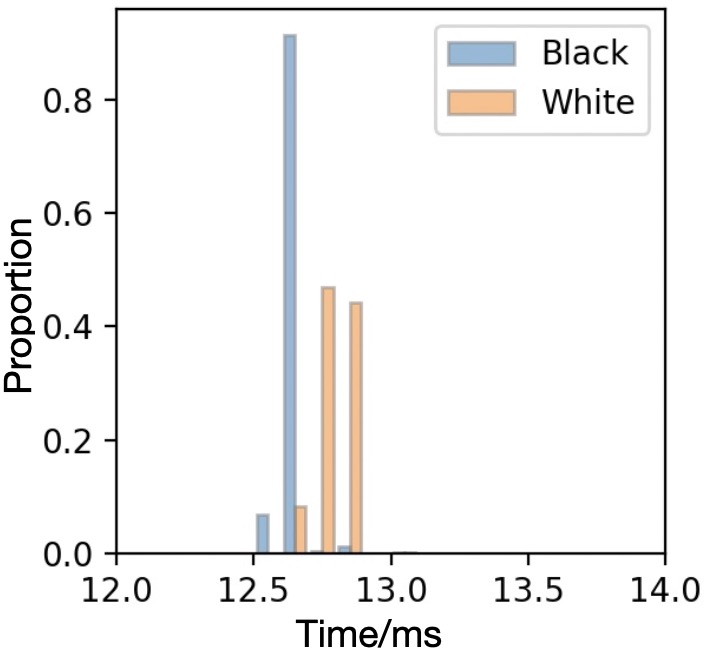}
   \caption{SLC access time for pixels}
   \label{fign51}
    \end{minipage}%
    \begin{minipage}{0.48\linewidth}
        \centering
        \includegraphics[width=0.9\textwidth]{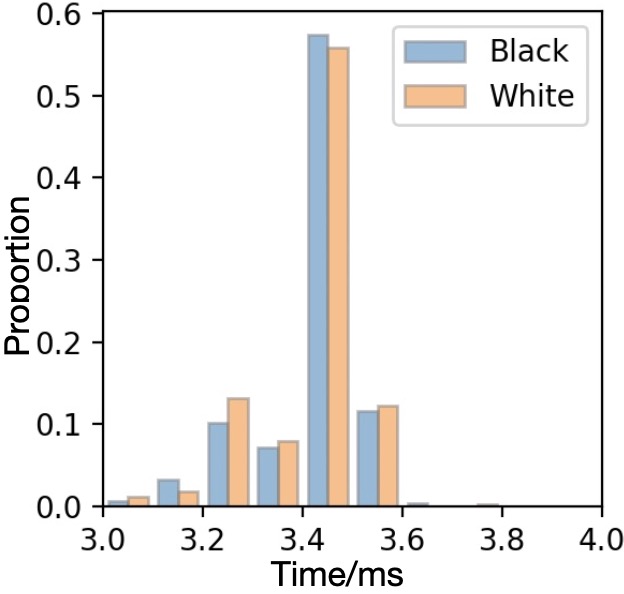}
   \caption{L2 access time for pixels}
   \label{fign52}
       \end{minipage}
\end{figure}

Figure~\ref{fign51} captures the distinct behavior of the SLC occupancy channel when processing black versus white pixels on Apple M1 chrome browser.
By choosing an appropriate threshold between these two distributions, we achieved a pixel recognition accuracy of 92\%.  
In contrast, Figure~\ref{fign52} presents the corresponding data for the L2 cache-channel.  It does not demonstrate significant differences between the two pixel colors, which underscores the limitations of the L2 cache occupancy channel for this type of pixel-stealing attack.

We further extended our experiments to an Apple M3-Pro SoC, which achieves similar accuracy.  We also experimented with different browsers.  Due to Safari’s lower timer precision (with a granularity of only 1 millisecond), we implemented a sweep-counting technique instead of directly measuring SLC access times. This technique counts how many times the buffer can be traversed within a fixed time window of 10 seconds. As demonstrated in Table \ref{tab:attack_comparison}, this alternative timing measurement method proves effective, though with a slightly lower accuracy compared to the direct SLC access time measurements for Chrome.

\begin{table}[htbp]
   \centering
   \caption{Pixel recognition accuracy in different settings}
   \label{tab:attack_comparison}
   \begin{tabular}{llcc}
       \hline
       \textbf{SoC} & \textbf{Browser} & \textbf{Attack Technique} & \textbf{Accuracy} \\
       \hline
       M1 & Chrome & SLC access time & 92\% \\
       M1 & Safari & Sweep-counting  & 84\% \\
       M3 Pro & Chrome & SLC access time & 94\% \\
       M3 Pro & Safari & Sweep-counting & 85\% \\
       \hline
   \end{tabular}
\end{table}

These results show that the SLC occupancy channel is uniquely effective in the pixel stealing attack, which is due to the SVG filter operations being GPU-bound and only affecting the SLC that is shared between CPU and GPU. \rev{Our approach achieves a pixel reading speed of 2 seconds per pixel in Chrome, making it 2.5 times faster than GPU-ZIP\cite{wang2024gpu}, the only prior pixel-stealing attack that remains unmitigated on Apple M1 devices. In contrast, the reading speed in Safari is slower, at 10 seconds per pixel.} These findings are consistent with the conclusions drawn in Sections \ref{sec3} and \ref{sec:sec4}, where we demonstrated that the SLC occupancy channel excels in a wide range of application scenarios, particularly those involving GPU-centric processes.

\subsection{Attack Cases}

\begin{figure}[htbp]
   \centering
   \begin{subfigure}[b]{0.1\textwidth}
       \centering
       \includegraphics[width=\textwidth]{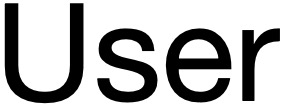}
       \caption{Original text}
       \label{fign53:sub1}
   \end{subfigure}
   \hfill
   \begin{subfigure}[b]{0.1\textwidth}
       \centering
       \includegraphics[width=\textwidth]{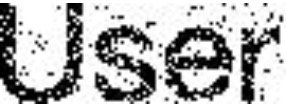}
       \caption{Retrieved text}
       \label{fign53:sub2}
   \end{subfigure}
   \hfill
   \begin{subfigure}[b]{0.1\textwidth}
       \centering
       \includegraphics[width=\textwidth]{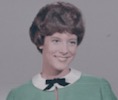}
       \caption{Original image}
       \label{fign53:sub3}
   \end{subfigure}
   \hfill
   \begin{subfigure}[b]{0.1\textwidth}
       \centering
       \includegraphics[width=\textwidth]{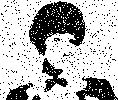}
       \caption{Retrieved image}
       \label{fign53:sub4}
   \end{subfigure}

   \caption{Pixel stealing attack}
   \label{fign53}
\end{figure}

To demonstrate the effectiveness and potential threats of our attack in practical scenarios, we simulated two attack cases - stealing text and image content from web pages. These attacks mirror common situations on many real websites, where usernames and profile pictures are typically displayed in a corner of the page.
Figure \ref{fign53} illustrates the original images and the reconstructed results from a Chrome browser running on an Apple M1 chip. The results clearly demonstrate that our attack successfully reconstructed both the text and image content. This outcome indicates that our attack method remains effective in stealing sensitive visual information, even on modern hardware and browsers.

The security implications of such attacks are significant. Many popular websites are allowed to be embedded via iframes on a third-party website  while displaying user identity information somewhere on the page, for example, Wikipedia and LinkedIn.
Attackers could exploit these opportunities with some malicious attack code embedded in their webpage and access the target webpages with iframes.
\rev
{Similar to the findings in Section 4.3, in real-world scenarios beyond the clean environment, most background applications and static browser tabs have negligible impact on the attack's effectiveness. However, other memory-intensive applications and webpages can prevent the attack by saturating the timing, thereby rendering the side-channel data noninformative.}

\section{Screen-capturing Attack}\label{sec6}
Building upon our foundational SLC occupancy channel and the pixel stealing attack demonstrated in Chapter 5, we next introduce a novel and more versatile attack that significantly expands the scope of potential threats. While the previous attack leveraged website iframes to magnify individual pixels to full-screen and monitor SVG filtering via the SLC occupancy channel to differentiate black and white pixels, our new screen-capturing attack transcends these limitations, presenting a more general and pervasive threat.

Our innovative approach begins with the key observation that even without SVG filters, significant differences in GPU memory usage exist when displaying large areas of black versus white on the screen, which can be detected by our SLC occupancy channel. Building on this insight, we developed a sophisticated algorithm capable of monitoring memory consumption during the rendering of each frame, capturing the dynamics of a frame's lifecycle. This advancement allows us to extract fine-grained information from specific screen regions without the need to artificially expand website iframes to full-screen size.

This screen-capturing attack specifically targets the rendering process on Apple M1 SoCs, leveraging the GPU's involvement in display operations. 
It can potentially extract information from any content rendered on the screen, including native applications, system UI elements, and full-screen applications, with an adequate monitoring granularity of 57 rows of pixels out of the screen's 1600 rows.

In the following sections, we first examine the sensitivity of the SLC occupancy channel to GPU activities during normal image rendering, then describe our approach to monitor individual frames with high precision. Finally, we demonstrate the effectiveness and broad applicability of this attack through two real-world scenarios: barcode retrieval and printed digit recovery, showcasing its potential to compromise sensitive information across various use cases beyond web-based contexts.

\subsection{Sensitivity of the SLC occupancy channel to GPU activities}\label{sec52}

In this section, we provide evidence supporting the relationship between SLC usage and the content displayed on the screen during normal image rendering, without relying on web-based techniques used in Chapter 5. We conduct experiments by displaying various images directly on the screen and monitoring the SLC usage through our SLC occupancy channel. This approach demonstrates that changes in displayed content significantly influence the SLC behavior. 
Our findings reveal that the GPU's memory usage patterns can be detected through the SLC for any on-screen content, providing a foundation for our subsequent screen-capturing attack that goes beyond web-based scenarios and expands the attack surface to all displayed information. 

\textbf{Experimental Setup:}
The experimental setup utilizes a MacBook Air with M1 SOC. We repeatedly profile using alternated order access pattern, counting SLC hits and calculating evicted cache lines, which serve as a proxy for GPU memory usage. \rev{Same as prior two attacks, all experiments in this section were conducted in a clean environment with no other applications running in the background.}
 
\begin{figure}[ht]
  \centering
  \begin{subfigure}[b]{0.11\textwidth}
    \includegraphics[width=\textwidth]{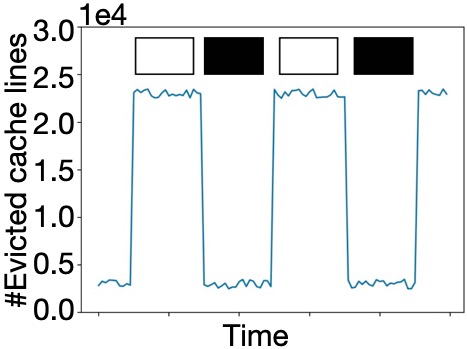}
    \caption{}
    \label{fig4bw}
  \end{subfigure}
  \hfill
  \centering
  \begin{subfigure}[b]{0.11\textwidth}
    \includegraphics[width=\textwidth]{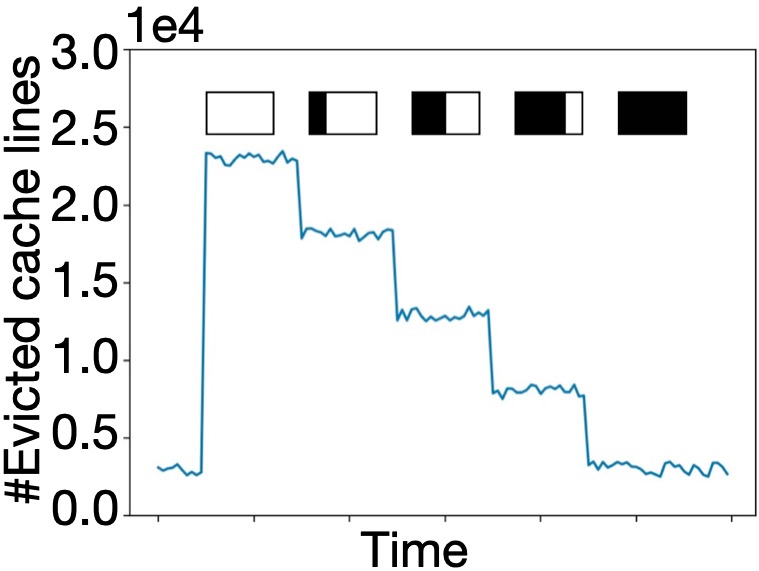}
    \caption{}
    \label{fig4jianjin}
  \end{subfigure}
  \hfill
  \begin{subfigure}[b]{0.11\textwidth}
    \includegraphics[width=\textwidth]{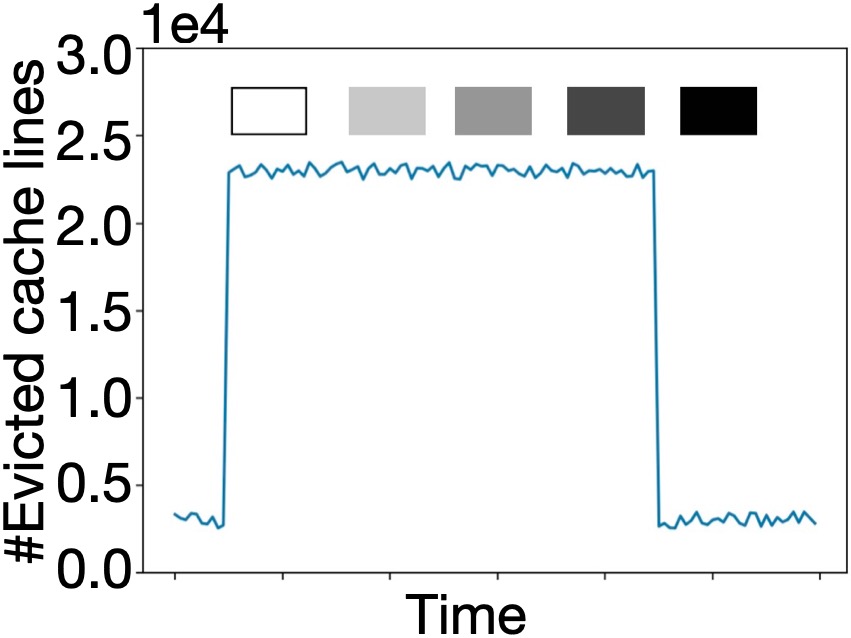}
    \caption{}
    \label{fig4g}
  \end{subfigure}
  \hfill
  \begin{subfigure}[b]{0.11\textwidth}
    \includegraphics[width=\textwidth]{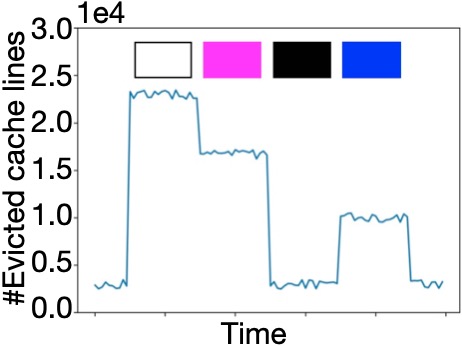}
    \caption{}
    \label{fig4cor}
  \end{subfigure}
  \label{fig4m}
  \caption{Evicted cache lines for different slides}
\end{figure}

We first show a sequence of slides alternating between complete darkness (all pixels black) and pure whiteness on the screen and use our SLC occupancy channel to measure the memory usage. We observe that the GPU's memory usage is significantly lower for displaying black pixels than white pixels, shown in Figure~\ref{fig4bw}.

Next, we display a sequences of slides with decreasing portion of white pixels, with the results show in Figure~\ref{fig4jianjin}. We observe a positive correlation between the memory usage and the percentage of white pixels. 
We then change the slides to be grayscale, and the results are shown in Figure~\ref{fig4g}. Interestingly there is no correlation between the memory usage and the grayscale level of the screen. 

We further analyze other colors, and discover that the GPU's memory usage is  decreased proportionally as the number of RGB color pixel values being zero increases, as shown in Figure~\ref{fig4cor}, where black is (0,0,0), pink is (0,255,255), blue is (0,0,255), and white is (255, 255, 255) in RGB format.

  Based on these results, we hypothesize that an increase in the number of zeros across screen pixels correlates with reduced GPU memory usage. This could be also due to the GPU's compression mechanism, which results in less memory usage and data transfer with more zero values.  
  Due to the lack of publicly available documentation, it is challenging to pinpoint the exact reason for this phenomenon. Nonetheless, understanding this correlation between zero pixel values and GPU memory usage is sufficient for us to develop a privacy-concerning screen display snooping attack.

  \subsection{Monitoring one frame}\label{sec53}
  To precisely measure GPU memory usage corresponding to screen displays, we monitor memory consumption during the rendering of one frame, aiming to capture dynamics during a frame's lifecycle. MacOS's Vsync synchronizes the screen refresh rate with graphical output at 60 FPS, with 16.7ms between Vsync signals. We utilize CVDisplayLink to capture Vsync signals and align the measurements.

\begin{figure}[t]
    \centering
    \includegraphics[width=0.8\columnwidth]{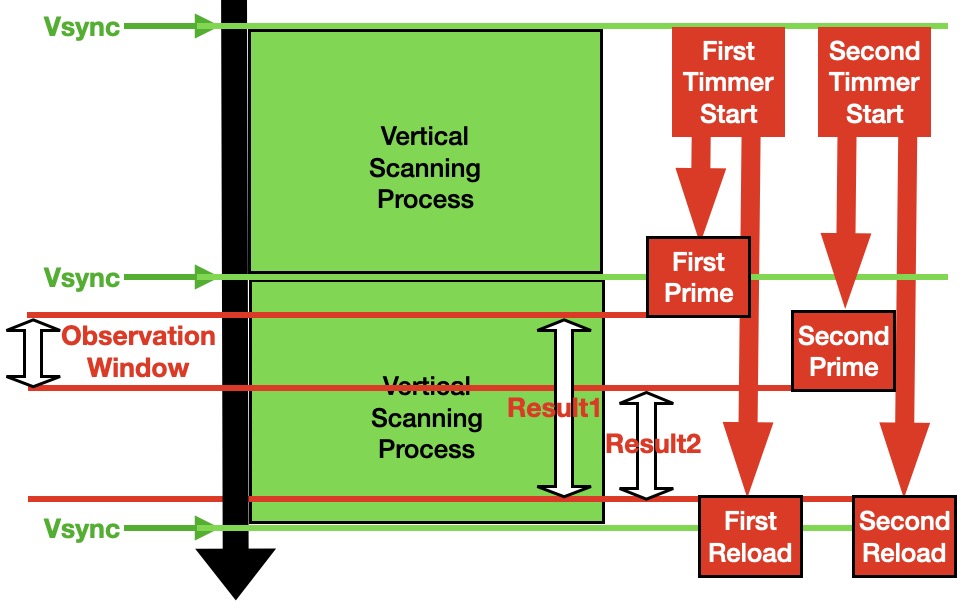}
    \caption{Precisely monitoring frame activities}
    \label{fig44}
    \vspace{-5mm}
\end{figure}

  Our monitoring framework, depicted in Figure~\ref{fig44}, uses a Prime \& Reload technique to capture the GPU's memory usage over very short time intervals. 
  The Prime and Reload are both profiling phases described in Section~\ref{SLC}, without or with measuring SLC hits.
  However, due to the inherent time consumption of the prime and reload processes themselves, the precision of measurements is compromised when these phases are closely spaced. 
  To address this challenge, we implement a strategy of dual-set of prime and reload, with deliberately controlled timing for each process, so that one single frame of screen rendering can be monitored in detail. 
      In this setup, both reloads start simultaneously, while the primes are initiated at different moments. By calculating the differential of the evicted cache lines measurements from these two prime and reload cycles, we can infer the memory usage within the time interval between the two primes. This short interval, which we define as an 'observation window,' allows us to analyze GPU memory usage over extremely brief periods, enhancing the granularity and accuracy of our measurements.

\begin{figure}[t]
    \centering
    \includegraphics[width=\columnwidth]{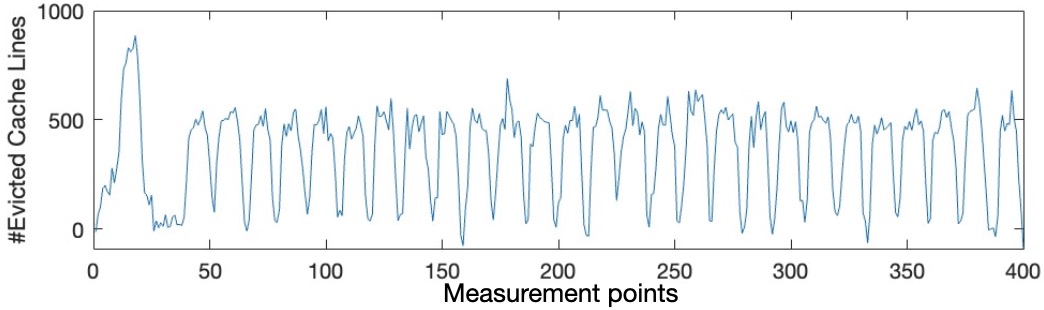}
    \caption{Trace of a full frame}
    \label{fig45}
    \vspace{-5mm}
\end{figure}

Figure~\ref{fig45} shows a trace for one frame, with a 0.4ms observation window (about $\frac{1}{40}$ frame) sliding with an increment of 0.04ms ($\frac{1}{400}$ frame). This trace, with 400 measurement points, shows that there are 28 short epochs during the frame rendering process, where the GPU memory usage is significantly active at the peak (we define as the flash point). The first epoch demonstrates a higher GPU memory usage compared to the subsequent epochs.

\begin{figure}[t]
    \centering
    \includegraphics[width=0.9\columnwidth]{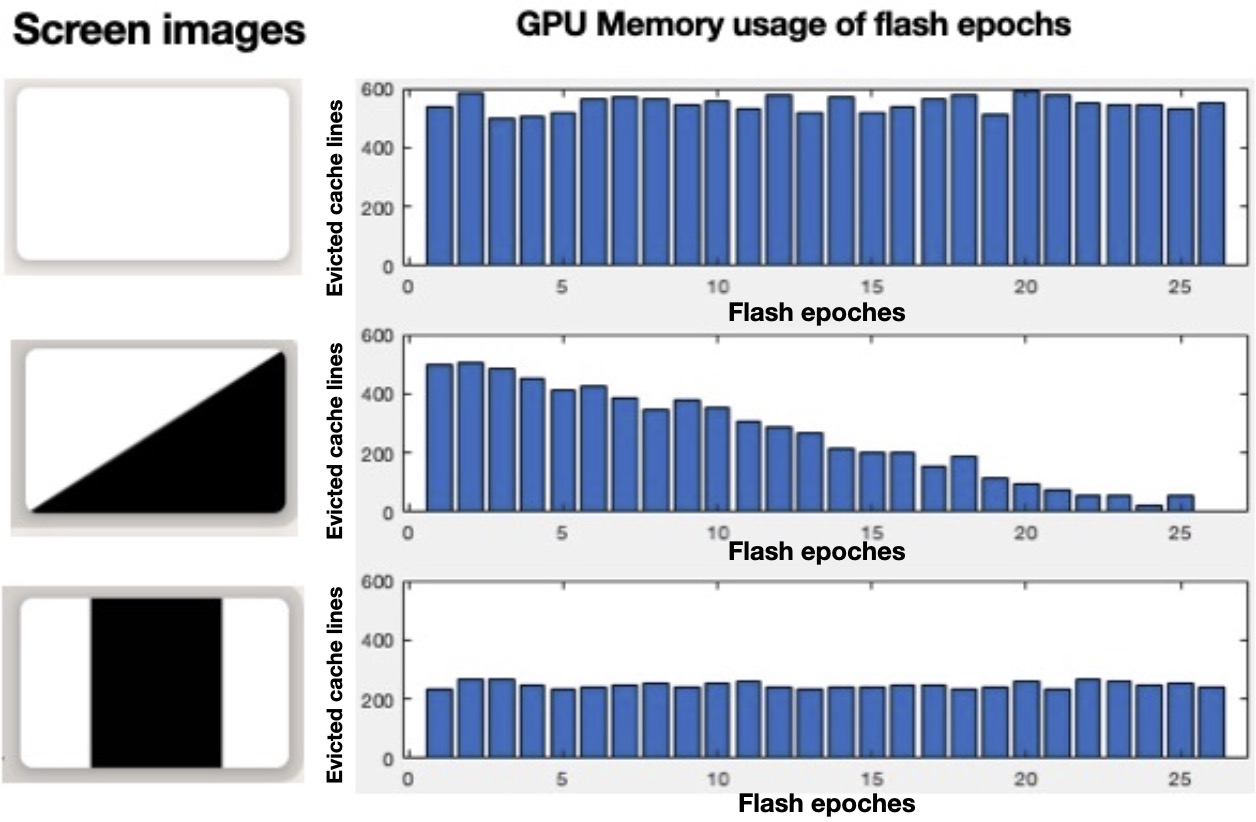}
    \caption{Correlation between a trace of flash points and the screen bars (with varying zero pixel values)}
    \label{fig47}
    \vspace{-5mm}
\end{figure}

Our experiments revealed that these epochs correspond to 28 evenly divided regions of the screen, vertically from top to bottom. As the screen is 2560*1600 pixels, each flash point can be mapped to a 2560*57 pixels region. We observe a high correlation between the memory usage at each flash point and the portion of pixel values at zero in its corresponding screen region (1/28 bar of the screen). Figure~\ref{fig47} shows several instances demonstrating this relationship. 
Note we excluded the first epoch for its peculiarity. The strong correlation observed suggests that this SLC occupancy channel can be exploited to extract more patterns from the screen.

\rev{However, detecting the screen content through flash point patterns has several limitations. First, the technique only works when the screen predominantly displays large blocks of solid colors (>90\% of screen area, with each block exceeding 256×256 pixels). The pattern detection becomes difficult when more than 10\% of the screen contains complex images with pixel-level variations. Moreover, the measurement process itself takes approximately 5 minutes and requires completely still screen - even mouse movements can disrupt the readings. Despite these limitations in capturing fine-grained screen details, we demonstrate in Section \ref{sec54} that this side-channel remains effective for specific screen-capturing attack scenarios.}

\subsection{Attack with pattern recognition}\label{sec54}

\begin{figure}[t]
  \centering
  \includegraphics[width=0.8\columnwidth]{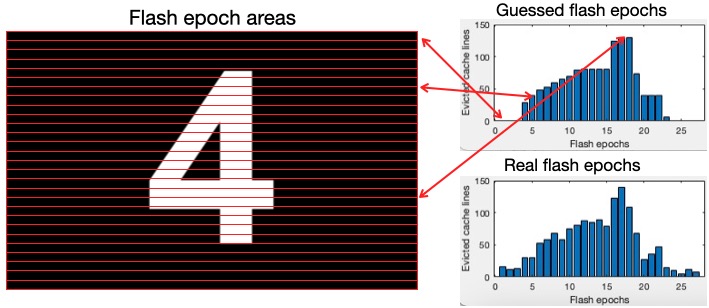}
  \caption{Screen display snooping attack method}
  \label{fig51}
  \vspace{-5mm}
\end{figure}

Building upon these insights, we propose a screen display snooping attack to recognize static images using pattern recognition. Assuming the adversary knows all potential images, they build a library of memory usage patterns by dividing each image into 28 bars and guessing flash point values based on white pixel proportions (Figure~\ref{fig51}). The attacker then obtains actual flash point values during screen rendering and compares them with the library to identify the best match. We apply this attack to barcode retrieval and printed digit recognition to demonstrate its practicality and effectiveness.

\subsubsection{\textbf{Barcode retrieving}}

Our first application of the screen display snooping attack targets identifying barcodes shown. We choose a classic barcode format, Interleaved 2 of 5 (ITF). This format is known for its efficiency in encoding numeric data, widely used in warehouse inventory, supply chain distribution, and various industrial applications. In ITF, each character is represented by a set of five bars, two of which are wide and three are narrow. A unique aspect of ITF is that two digits are encoded in each character, making it a compact and dense representation. The barcode begins and ends with specific start and stop patterns, ensuring correct orientation and readability. According to the standard ITF barcoding, the width of the wide bars is three times that of the narrow bars.

\begin{figure}[t]
  \centering
  \begin{subfigure}[b]{0.5\columnwidth}
    \includegraphics[width=0.9\columnwidth]{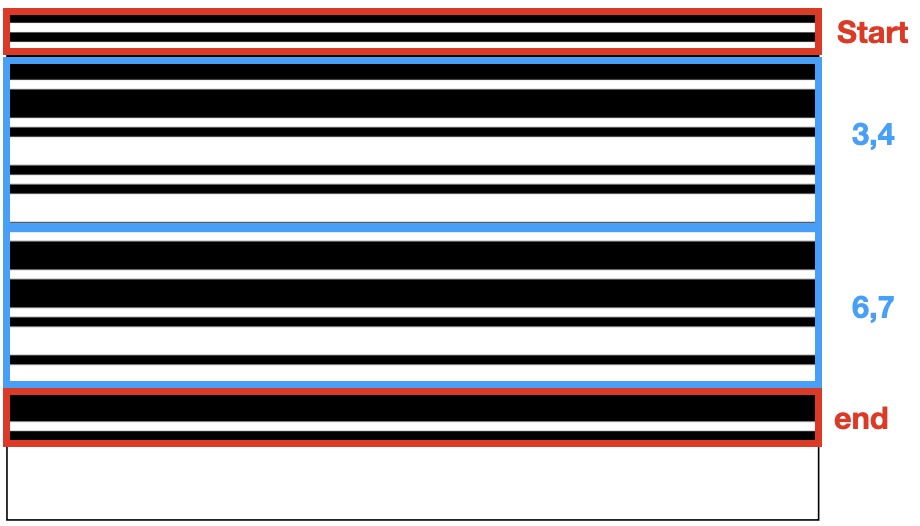}
      \caption{Characters}
      \label{fig52:a}%
  \end{subfigure}%
  \hfill
  \begin{subfigure}[b]{0.5\columnwidth}
    \includegraphics[width=0.8\columnwidth]{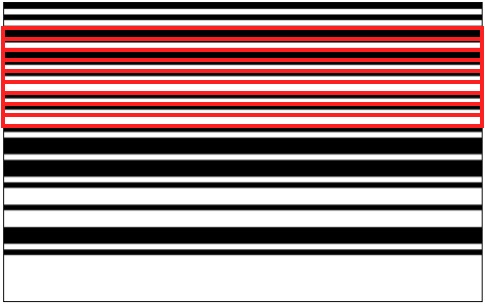}
      \caption{Flash epoch segments}
      \label{fig52:b}
  \end{subfigure}
  \vspace{-5mm}
  \caption{ITF barcode}
  \label{fig52}%
  \vspace{-5mm}
\end{figure}

An example of an ITF barcode with a narrow bar height of 30 pixels is shown in Figure~\ref{fig52:a}. This barcode contains two characters, each encoding two digits. 
For the first character, its area on the screen corresponds to nine flash epochs, as depicted in Figure~\ref{fig52:b}. During our pattern generation phase, we calculate the expected memory usage for these nine flash points for all possible two-digit values that the character can represent, ranging from 00 to 99. To determine the actual value of the character, we collect a trace of flash points from the screen rendering and compare it against the pre-computed patterns. The pattern that yields the smallest distance, measured by Mean Squared Error (MSE), is identified as the correct value for the character. In the example shown, the first character is correctly recognized as ``34.''

\begin{figure}[t]
  \centering
  \includegraphics[width=0.8\columnwidth]{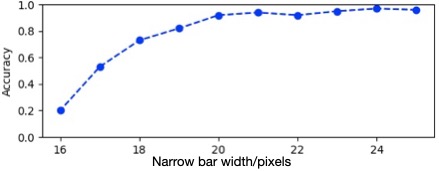}
  \caption{Attack accuracy vs. the narrower bar width}
  \label{fig54}
  \vspace*{-5mm}
\end{figure}

To evaluate the effectiveness of our attack on ITF barcodes, we tested the recognition accuracy for different narrow bar heights. The results are presented in Figure~\ref{fig54}, which demonstrates that our attack achieves a 90\% accuracy rate when the narrow bar height is greater than or equal to 20 pixels. Considering the height limitation of the screen, this implies that our attack can handle ITF barcodes containing up to 8 digits (i.e., 4 characters) in a full-screen setting. 
 
\subsubsection{\textbf{Print digits recovery}}

\begin{wrapfigure}{r}{0.4\columnwidth}
  \vspace{-4mm}
  \hspace{-3mm}
   \begin{minipage}[t]{0.4\columnwidth} 
     \centering
     \captionof{table}{\small Confusion matrix for single-digit recognition} 
     \label{tab:confusion_matrix}
        \vspace{-1mm}
     \setlength{\tabcolsep}{1pt}
     \resizebox{0.9\columnwidth}{!}{%
       \begin{tabular}{c|cccccccccc}
         & 0 & 1 & 2 & 3 & 4 & 5 & 6 & 7 & 8 & 9 \\
         \hline
         0 & 89 & 0 & 0 & 0 & 0 & 0 & 3 & 0 & 5 & 3 \\
         1 & 0 & 100 & 0 & 0 & 0 & 0 & 0 & 0 & 0 & 0 \\
         2 & 0 & 0 & 96 & 3 & 0 & 0 & 0 & 0 & 1 & 0 \\
         3 & 0 & 0 & 1 & 98 & 0 & 1 & 0 & 0 & 0 & 0 \\
         4 & 0 & 0 & 0 & 0 & 100 & 0 & 0 & 0 & 0 & 0 \\
         5 & 0 & 0 & 0 & 2 & 0 & 97 & 1 & 0 & 0 & 0 \\
         6 & 3 & 0 & 0 & 0 & 0 & 0 & 89 & 0 & 5 & 3 \\
         7 & 0 & 0 & 0 & 0 & 0 & 0 & 0 & 100 & 0 & 0 \\
         8 & 10 & 0 & 0 & 0 & 0 & 0 & 5 & 0 & 74 & 11 \\
         9 & 3 & 0 & 0 & 0 & 0 & 0 & 2 & 0 & 7 & 88 \\
       \end{tabular}
     }
   \end{minipage}%
   \end{wrapfigure}

The second application of our screen display snooping attack is print digits displayed. The victim displays a series of Arabic numerals in white on a black background full-screen, as shown in Figure~\ref{fig51}. Such digits may contain sensitive information, such as verification codes and passwords, that the attacker aims to steal. Initially focusing on a single digit, the results of this attack are presented in a confusion matrix, revealing an accuracy rate of 91\%. 
Most digits exhibit high recognition accuracy, indicating the effectiveness of the screen display snooping attack in identifying individual numerals. However, it is observed that certain pairs of visually similar digits, such as 0 and 8, have a higher probability of being misclassified as each other. This can be attributed to their resemblance in shape and the shared visual features that may lead to similar SLC usage patterns.

  We further investigate scenarios where two and three digits are displayed simultaneously on the screen. In these cases, our attack is unable to distinguish the order of the digits. For example, the digit pairs 81 and 18, or the digit triads 119 and 911, would have the same guessed flash points. When calculating the accuracy, we consider different orderings of the same digits as a single label. This means that when the attacker succeeds, they can only determine the count of each digit present, but not their order. The accuracies for these cases are detailed in Table~\ref{tab:accuracy}.

  \begin{wrapfigure}{r}{0.4\columnwidth}
     \vspace{-5mm}
     \begin{minipage}[t]{0.4\columnwidth} 
       \centering
       \captionof{table}{\small Recognition accuracy for multiple digits}
       \label{tab:accuracy}
       \vspace{-1mm}
       \setlength{\tabcolsep}{3pt}
       \resizebox{\columnwidth}{!}{%
       \begin{tabular}{c|ccc}
         \hline
         \textbf{Digit Count} & \textbf{Top-1} & \textbf{Top-5} & \textbf{Top-10} \\
         \hline
         Single-digit & 91.2\% & 100\% & 100\% \\
         Two-digit & 52.9\% & 63.0\% & 63.9\% \\
         Three-digit & 21.4\% & 27.2\% & 28.2\% \\
         \hline
       \end{tabular}
       }
     \end{minipage}
     \vspace{-5mm}
    \end{wrapfigure}

  The accuracy decreases significantly as the number of digits increases. This can be attributed to the increased precision required to differentiate between the possible combinations of digits. With each additional digit, the number of possible combinations grows by a factor of 10, while the noise level remains constant. However, the attack still poses a threat by narrowing down possible values, as sensitive information like verification codes or PINs often consists of a limited number of digits.

\section{Countermeasure against Cache Occupancy Attacks}
In this section, we propose a novel countermeasure against SLC cache occupancy attacks, building upon and significantly enhancing the cache masking technique introduced by Oren et al.~\cite{shusterman2020website}. While their approach showed only moderate effectiveness, reducing attack accuracy by a mere 6\%, our method addresses the fundamental limitations that hindered its performance.

Our analysis in Section~\ref{sec3} revealed that the ineffectiveness of prior techniques stemmed from the uneven distribution of the mask buffer across L2 cache sets. This allowed attackers to exploit overlooked cache sets, maintaining the viability of their side-channels. Our enhanced approach ensures a more comprehensive coverage of cache sets, effectively disrupting the entire cache occupancy channel.

Unlike subsequent works~\cite{li2022cache,seonghun2023defweb} that primarily focused on adding noise to the cache occupancy channel, our countermeasure fundamentally alters the relationship between victim activities and observable cache states. By leveraging our deep understanding of the SLC's structure and behavior, we've developed a method that targets the SLC occupancy channel itself, rather than specific attacks or applications. This ensures no discernible impact of victim activity is visible to potential attackers. 

In the following subsection, we detail our proposed countermeasure, its implementation, and demonstrate its effectiveness against various attack scenarios, including high-resolution attacks.

\subsection{Proposed countermeasure}
We propose two enhancements to the cache masking process to address the issue identified.  First, we allocate a new buffer for each iteration of the loop for cache masking and release the buffer at the end. This approach ensures that the evicted cache lines may vary in each iteration. 
It becomes more challenging to consistently miss a particular cache set throughout the entire loop.
Second, we increase the buffer size used for mitigation to exceed the L2 cache size. By doing so, each cache set has a higher probability of being completely evicted.

\begin{figure}[ht]
    \centering
    \begin{subfigure}[b]{0.48\columnwidth}
    \includegraphics[width=\linewidth]{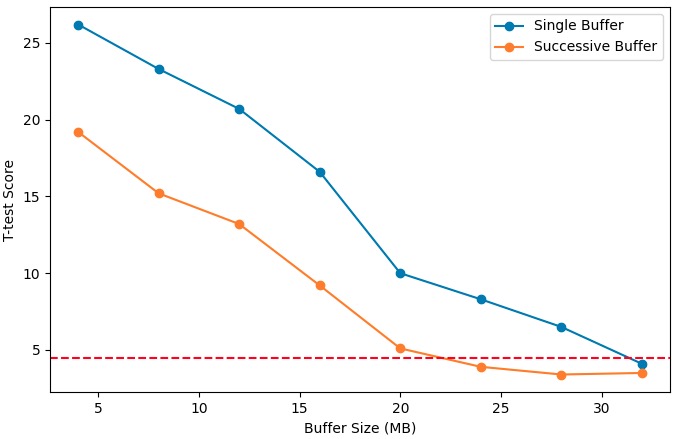}
    \caption{\scriptsize Against L2 occupancy channel}
    \label{fig61}
    \end{subfigure}
    \hfill
    \begin{subfigure}[b]{0.48\columnwidth}
\includegraphics[width=\linewidth]{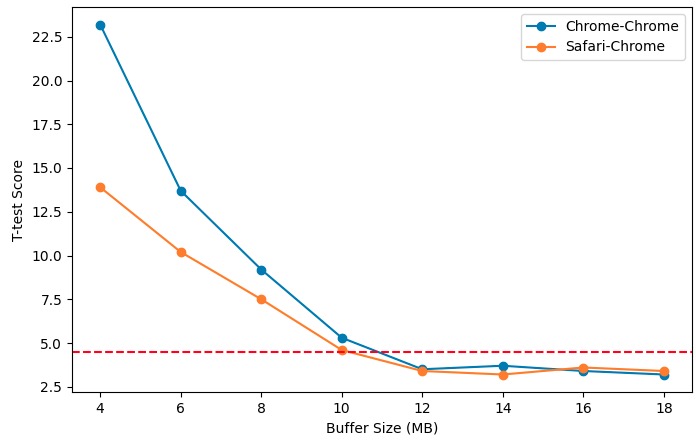}
\caption{\scriptsize Against SLC occupancy channel}
\label{fig63}
\end{subfigure}
 \caption{Effectiveness of the proposed countermeasures}
        \label{countermeasures}%
    \end{figure}

We employ the benchmark introduced in Section~\ref*{vali} to evaluate the impact of the mitigation on the previous L2 cache occupancy channel. 
 We test  with the spy and victim both operating in Chrome (cross-tab scenario). Figure~\ref{fig61} shows the effectiveness of our countermeasure under different buffer sizes. We evaluate two cases: using a single buffer (without the first enhancement) and continuously allocating and releasing a new buffer in each iteration (with the first enhancement). 
The results show for the single-buffer case, when the buffer size is larger than 32 MB (almost three times the L2 cache size), the cache occupancy side-channel is ineffective. However, for the new-buffer case, the buffer size threshold is 22MB. 

However, when applying this countermeasure to our SLC occupancy side-channel, it fails to effectively mitigate the SLC channel. 
We hypothesize that although this method can fill the SLC by evicting L2, the filling rate is not sufficiently fast.
To address this issue, we propose a specific SLC masking scheme. The main change is setting the stride of accessing the buffer to be 8 KB, which is significantly larger than the previous stride of 128 bytes used in the previous masking approach.
\rev{This large stride allows the masking buffer to bypass the L2 cache and directly fill the SLC, making it effective against SLC occupancy channels. However, this design choice means that our SLC masking scheme has minimal impact on L2 cache occupancy - since the masking buffer largely bypasses L2, it cannot effectively mitigate L2-based occupancy channels. }
Figure~\ref{fig63} present the effectiveness of the new SLC masking scheme. 
The results demonstrate that when the buffer size is greater than or equal to 12MB, the SLC masking scheme counteract against the SLC occupancy side-channel effectively.

\subsection{Performance evaluation}\label{eva}
\begin{figure}[ht]
\centering
\includegraphics[width=0.8\columnwidth]{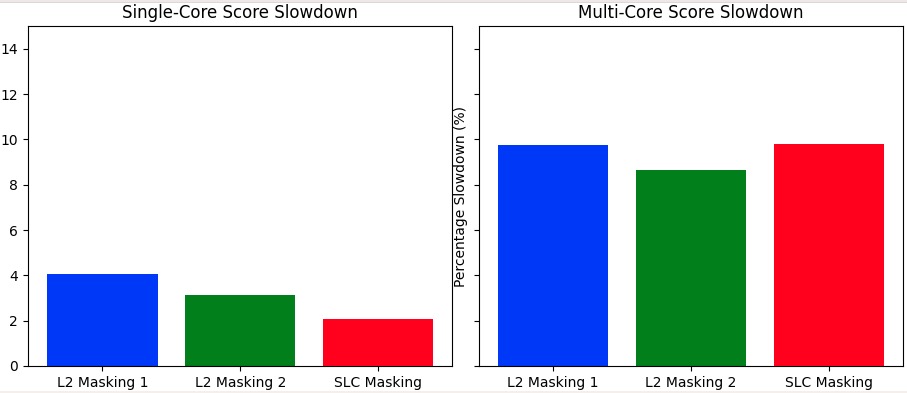}
\caption{Performance impact of mitigations}
\label{fig:perf}
\end{figure}

To evaluate the impact of cache masking techniques on the system performance, we employ Geekbench\cite{geekbench} to measure the influence of two L2 cache masking implementations (L2 masking 1: new-buffer at size 24MB; L2 masking 2: single-buffer with a buffer size=32MB) and the SLC masking implementation (stride=8KB and buffer size=12MB) on the performance of single-core and multi-core tasks. Figure~\ref{fig:perf} presents the evaluation results. 
Overall, the performance degradation caused by the three mitigation methods is relatively small, with single-core performance degradation by less than 5\% and multi-core degradation by less than 10\%.
We observe that the new-buffer implementation (L2 masking 1) has a more significant impact on the system performance, which suggests that the benefits of enhancement 1 are not sufficient to compensate for the overhead it introduces. 
The SLC masking implementation has a minor impact on single-core tasks but a more substantial effect on multi-core performance, which aligns with the role of the SLC in the system.

\section{Conclusion}

In this paper, we presented a comprehensive study of cache occupancy attacks targeting the System-Level Cache (SLC) in Apple M-series SoCs, revealing critical vulnerabilities in these modern heterogeneous computing systems. Our work makes several significant contributions to the field of hardware security:

\begin{enumerate}
    \item \textbf{Novel SLC Occupancy Channel:} We developed the first cache occupancy attack that effectively exploits the exclusive nature of the SLC, a feature previously unexplored in such attacks. By reverse-engineering the SLC's structure and sharing mechanisms, we uncovered its unique properties and devised a method to directly monitor its occupancy, bypassing lower-level caches.

    \item \textbf{Enhanced Attack Scenarios:} We demonstrated the superiority of our SLC occupancy channel over previous cache occupancy attacks, particularly in inter-cluster and CPU-GPU scenarios. This advancement significantly expands the attack surface in heterogeneous computing environments.

    \item \textbf{Versatile Attack Applications:} We showcased the effectiveness of our approach through three distinct attacks:
    \begin{itemize}
        \item A website fingerprinting attack that achieves high accuracy across various scenarios, including cross-browser setups.
        \item A cross-origin pixel stealing attack that exploits GPU compression characteristics to infer pixel values, overcoming existing security measures.
        \item A novel screen-capturing attack that can extract information from any on-screen content, representing a substantial escalation in potential privacy violations.
    \end{itemize}

    \item \textbf{Countermeasures:} We proposed and evaluated enhanced cache masking techniques to mitigate these attacks, addressing the shortcomings of previous countermeasures.
\end{enumerate}

Our findings highlight the urgent need for a reevaluation of security measures in modern SoC designs, particularly those employing heterogeneous architectures. The ability to exploit the SLC for cross-component attacks poses a significant threat to system-wide security and privacy.

This research demonstrates that as computing systems continue to evolve towards more complex and integrated architectures, understanding and mitigating these subtle yet powerful side-channel vulnerabilities becomes increasingly crucial. Our work serves as a foundation for securing heterogeneous computing environments against sophisticated microarchitectural attacks.

\begin{acks}
This work was supported in part by National Science Foundation under grants SaTC-1929300 and CNS-1916762. 
\end{acks}

\newpage
\bibliographystyle{ACM-Reference-Format}
\bibliography{refs}

\appendix
\section*{Appendix}
\section{SLC reverse engineering}\label{AppendixA}
\subsection{Reverse engineering Inclusiveness Policy}\label{InclusivenessPolicy}
Our experiments reveal that the SLC in Apple M1 employs a hybrid inclusiveness policy: inclusive with respect to the GPU cache but exclusive with respect to the CPU cache. To reach this conclusion, we conducted separate experiments to explore the relationship between the GPU cache and the SLC, as well as between the CPU cache and the SLC. 

\begin{figure}[t]
        \centering
        \begin{minipage}[t]{0.45\columnwidth}
            \centering
            \includegraphics[width=0.98\linewidth]{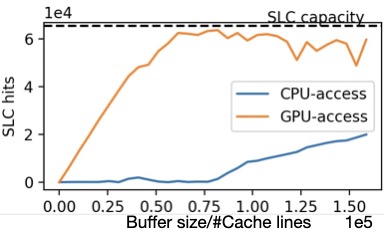}
            \caption{SLC capacities}
            \label{fign32}
        \end{minipage}
       \hfill
       \begin{minipage}[t]{0.45\columnwidth}
            \centering
            \includegraphics[width=0.98\linewidth]{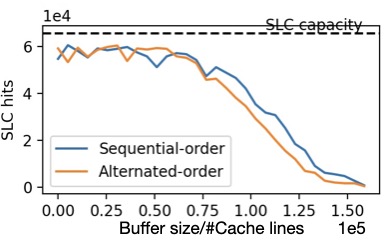}
            \caption{SLC eviction patterns}
            \label{fign31}
        \end{minipage}%
     \end{figure}

\noindent\textbf{GPU Cache and SLC Relationship:}
We first explored the relationship between the GPU cache and the SLC. We performed an experiment similar to that shown in Figure~\ref{fig32:b}, but from the GPU side and measure the number of SLC hits. 
The results, shown in Figure~\ref{fign32}, demonstrate a striking difference from CPU access patterns. When the GPU accesses the buffer, the number of SLC hits increases linearly with the buffer size until the buffer size reaches the SLC size. This behavior strongly suggests that the GPU cache is \textit{inclusive} with respect to the SLC, meaning that data loaded into the GPU cache is also present in the SLC.

\noindent\textbf{CPU Cache and SLC Relationship:}
    On the CPU side, in Apple M1, as the SLC (8~MB) is smaller than the L2 cache of the performance cores (12~MB), an inclusive policy between them is not likely. However, whether the SLC is \textit{non-inclusive} or \textit{exclusive} with respect to the CPU caches remains unclear. The key difference between non-inclusive and exclusive caches lies in what happens to the data in the SLC when it is accessed by the CPU. In an exclusive cache, when data is loaded from the SLC into the CPU's cache hierarchy, the corresponding data in the SLC is immediately invalidated. In contrast, a non-inclusive SLC keeps a backup copy of the data in it until it is evicted later by other activities.

    \rev{
    To infer the inclusiveness policy between the SLC and the CPU cache, we designed an experiment outlined in Algorithm 1. The algorithm systematically varies the buffer1 size and measures SLC hit counts to determine the cache behavior.}

\begin{algorithm}[h]
    \caption{SLC Inclusiveness Policy Test}\label{alg1}
    \begin{algorithmic}[1]
        \STATE $Buffer_{2}\_size \gets SLC\_size$
        \FOR{$Buffer_{1}\_size \gets 1$ to $1.6 \times 10^5$}
            \STATE $counter \gets 0$
            \begin{tikzpicture}[remember picture, overlay]
                \draw[decorate, decoration={brace, amplitude=10pt}, thick]
                    ([xshift=8em]0,-0.1) -- ([xshift=8em]0,-1.1)
                    node[midway, right=1em] {\textcircled{1}};
            \end{tikzpicture}
            \FOR{$i \gets 1$ to $Buffer_{1}\_size$}
                \STATE CPU accesses $Buffer_{1}[i]$
            \ENDFOR
            \begin{tikzpicture}[remember picture, overlay]
                \draw[decorate, decoration={brace, amplitude=8pt}, thick]
                    ([xshift=8em]0,-0.3) -- ([xshift=8em]0,-0.3)
                    node[midway, right=1em] {\textcircled{2}};
            \end{tikzpicture}
            \STATE GPU accesses $Buffer_{2}$
            \begin{tikzpicture}[remember picture, overlay]
                \draw[decorate, decoration={brace, amplitude=8pt}, thick]
                    ([xshift=6em]0,-0.3) -- ([xshift=6em]0,-3.6)
                    node[midway, right=1em] {\textcircled{3}};
            \end{tikzpicture}
            \IF{Sequential-order}
                \FOR{$i \gets 1$ to $Buffer_{1}\_size$}
                    \STATE CPU accesses $Buffer_{1}[i]$
                \ENDFOR
            \ENDIF
            \IF{Alternated-order}
                \FOR{$i \gets Buffer_{1}\_size$ to $1$}
                    \STATE CPU accesses $Buffer_{1}[i]$
                \ENDFOR
            \ENDIF
            \begin{tikzpicture}[remember picture, overlay]
                \draw[decorate, decoration={brace, amplitude=8pt}, thick]
                    ([xshift=17em]0,-0.1) -- ([xshift=17em]0,-2.3)
                    node[midway, right=1em] {\textcircled{4}};
            \end{tikzpicture}
            \FOR{$i \gets 1$ to $Buffer_{2}\_size$}
                \STATE $t \gets$ measure\_time(\text{CPU access } $Buffer_{2}[i]$)
                \IF{is\_SLC\_hit($t$)}
                    \STATE $counter \gets counter + 1$
                \ENDIF
            \ENDFOR
            \STATE store\_result($Buffer_{1}\_size$, $counter$)
        \ENDFOR
    \end{algorithmic}
\end{algorithm}

    \rev{
    There are four steps as shown in Algorithm~\ref{alg1}: CPU loads buffer1; GPU loads buffer2: CPU loads buffer1; CPU loads buffer2 and measure the SLC hit counts.        For the second load of \textit{buffer1}, we employ the \textit{sequential-order} and \textit{alternated-order} access patterns described earlier, respectively, and see if there are differences in the SLC hit counts.
  When the size of \textit{buffer1} exceeds the capacity of the L2 cache, step \textcircled{1} results in some earlier buffer1 data occupying the SLC (denoted as $buffer1_{SLC}$ and later buffer1 data stays in L2 (denoted as $buffer1_{L2}$) due to cache capacity conflict. Step \textcircled{2} evicts $buffer1_{SLC}$ from SLC. For the alternate-order access pattern, for step \textcircled{3}, the later buffer1 data is accessed first with L2 cache hit, and when earlier data is accessed, they have to be loaded from the memory to L2 cache which evicts some $buffer1_{L2}$ to SLC, at the same amount as $buffer1_{SLC}$. For the sequential-order access pattern, for step \textcircled{3}, the self-eviction effect, as previously described in Section~\ref{L2}, will take place and some $buffer1_{L2}$ content will be evicting from L2 to SLC first and later loaded to L2 again, at the same amount as $buffer1_{SLC}$ .  }

    Theoretically, if the SLC were non-inclusive with respect to the CPU cache, we would expect a significant difference in the number of SLC lines occupied by \textit{buffer1} data between the two access patterns, similar to the difference in SLC hits observed in Figure~\ref{accesspattern}. However, our experimental results, shown in Figure~\ref{fign31}, indicate that the difference between the two access patterns is minimal. This observation aligns with the behavior of an exclusive cache. Under an exclusive policy, when some $buffer1_{L2}$ is reloaded into the L2 cache during step \textcircled{3}, the corresponding data in the SLC is immediately invalidated to receive later evicted data from the L2 cache, resulting in total amount of  $buffer1_{SLC}$ .  This process does not displace other SLC lines, explaining the minimal difference observed between the two access patterns.

\subsection{Reverse engineering Set Index Mapping}\label{SetindexPolicy}
\begin{figure}[t]
        \centering
        \begin{subfigure}[b]{0.5\columnwidth}
            \includegraphics[width=\linewidth]{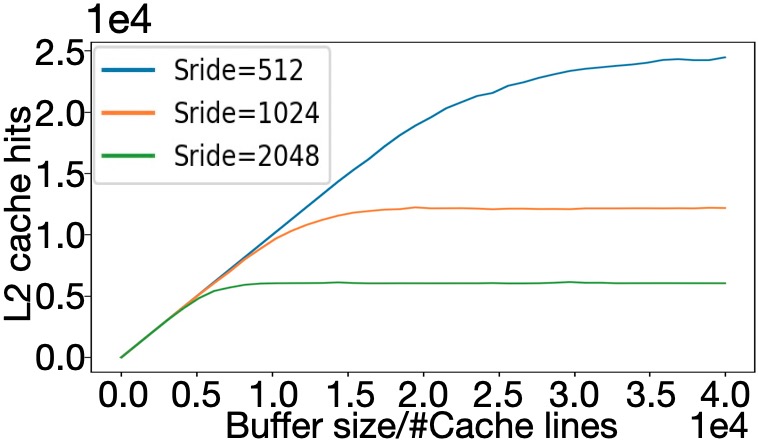}
            \caption{L2 utilization}
            \label{fig33:a}
        \end{subfigure}%
        \hfill
        \begin{subfigure}[b]{0.5\columnwidth}
            \includegraphics[width=\linewidth]{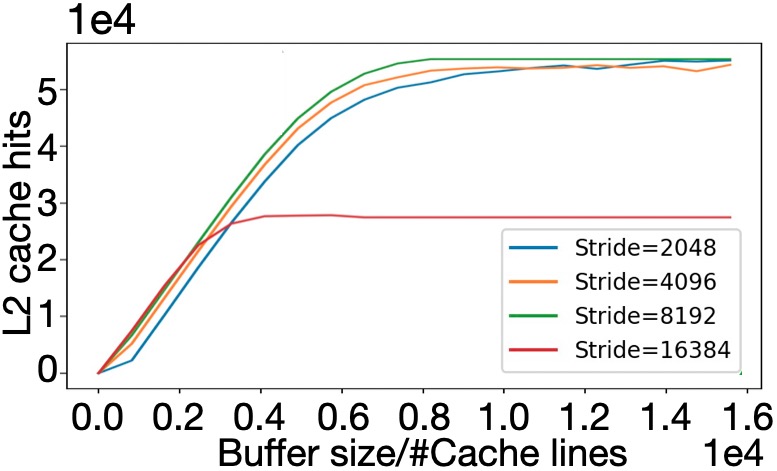}
            \caption{SLC utilization}
            \label{fig33:b}
        \end{subfigure}
        \caption{Cache utilization with different strides}
        \label{fig33}%
    \end{figure}

Our analysis reveals that the M1’s SLC excludes the lowest 13 bits of the physical address for indexing, instead utilizing bits starting from the 14th position and above. To verify this, we conducted experiments by fixing specific bits in the address and measuring the resulting SLC utilization. Figure~\ref{fig33} demonstrates how the utilization rates of the L2 and SLC caches change under different access patterns.

    In this experiment, we sequentially load a buffer with addresses using a defined stride, and measure the number of cache hits as the buffer size varies. For each buffer configuration, we record the number of cache hits on both the L2 and SLC caches, respectively, as depicted on the Y-axis. The experiment was conducted using an alternated-order access sequence with the chosen address stride. Similar to Figure~\ref{fig32:b}, as the buffer size increases, the cache hits for both L2 and SLC approach an upper limit, indicating cache utilization. For the L2 cache, the utilization is inversely proportional to the stride size due to the fixed stride causing some of the lowest bits in the address to remain constant. For example, a stride of 512 results in the lowest 9 bits being identical, while a stride of 1024 leads to the lowest 10 bits remaining the same. Since the L2 cache uses these fixed lowest bits to index sets, this pattern restricts the L2 cache sets that are utilized. 
    In contrast, the upper bound of SLC's cache hits remains close to about 90\% of the SLC's capacity, indicating that its utilization is not affected by the stride size when it's less than or equal to 8192 bytes. This finding suggests that the SLC does not utilize the lowest 13 bits for indexing cache sets. The bound of SLC's cache hits only begins to halve and becomes inversely proportional to the stride size when the stride is greater than or equal to 16384. This observation indicates that the SLC only employs the 14th bit and above for set indexing.

\subsection{Reverse engineering Replacement Policy}\label{ReplacementPolicy}

To provide further insight into the replacement policy of the SLC, we conducted an experiment with two buffers of equal size, accessed sequentially. We used a coarse-grained load+probe process. In the load phase, buffer 1 is accessed first, followed by buffer 2. In the probe phase, buffer 2 is accessed first, followed by buffer 1, and the loading times of both buffers are measured.
We also varied the buffer sizes for this experiment.

\begin{figure}[H]
    \centering
    \begin{subfigure}[b]{0.5\columnwidth}
        \includegraphics[width=\linewidth]{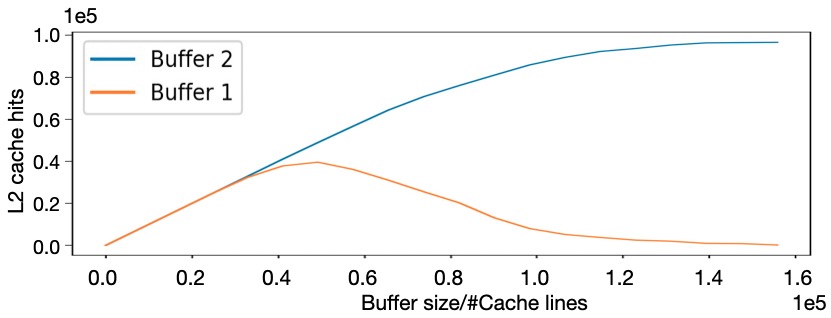}
        \caption{L2 hits}
        \label{fig34:a}
    \end{subfigure}%
    \hfill
    \begin{subfigure}[b]{0.5\columnwidth}
        \includegraphics[width=\linewidth]{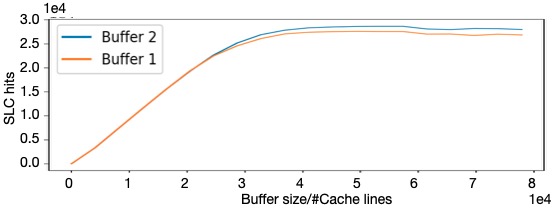}
        \caption{SLC hits}
        \label{fig34:b}
    \end{subfigure}
    \caption{Cache hits when accessing two buffers}
    \label{fig34}%
\end{figure}

Figure~\ref{fig34} illustrates the number of L2 and SLC cache hits for buffer 1 and buffer 2. For the L2 cache, which follows an LRU (Least Recently Used) replacement policy under single-core conditions, we observe that when the combined size of the two buffers approaches the L2 capacity, the cache hits for buffer 1 decrease significantly, while the hits for buffer 2 continue to increase. This confirms the expected LRU behavior, where the earlier accessed buffer is evicted first.

In contrast, the SLC cache hits for both buffers remain nearly identical across all buffer sizes, suggesting that the SLC replacement policy is independent of the access order, indicating a pseudo-random replacement policy. This makes it difficult for a single buffer to completely occupy the SLC, as the cache line replacement is not purely based on access recency but likely involves a form of randomization or pseudo-random selection. This is why, as shown in Figure~\ref{fig34:b}, the SLC utilization only reaches about 90\%.

\section{Websites Included in Datasets}\label{Appendix_B}

\noindent
\begin{tabbing}
    \hspace{3cm} \= \hspace{7.5cm} \= \kill
    \rev{www.google.com} \> \rev{www.youtube.com} \\
    \rev{www.reddit.com} \> \rev{www.facebook.com} \\
    \rev{www.amazon.com} \> \rev{www.pornhub.com} \\
    \rev{www.wikipedia.org} \> \rev{www.yahoo.com} \\
    \rev{www.duckduckgo.com} \> \rev{www.twitter.com} \\
    \rev{www.weather.com} \> \rev{www.xvideos.com} \\
    \rev{www.instagram.com} \> \rev{www.fandom.com} \\
    \rev{www.bing.com} \> \rev{www.cnn.com} \\
    \rev{www.tiktok.com} \> \rev{www.espn.com} \\
    \rev{www.nytimes.com} \> \rev{www.xnxx.com} \\
    \rev{www.9gag.com} \> \rev{www.foxnews.com} \\
    \rev{www.quora.com} \> \rev{www.ebay.com} \\
    \rev{www.linkedin.com} \> \rev{www.imdb.com} \\
    \rev{www.office.com} \> \rev{www.twitch.tv} \\
    \rev{www.xhamster.com} \> \rev{www.openai.com} \\
    \rev{www.live.com} \> \rev{www.microsoft.com} \\
    \rev{www.walmart.com} \> \rev{www.accuweather.com} \\
    \rev{www.onlyfans.com} \> \rev{www.usps.com} \\
    \rev{www.netflix.com} \> \rev{www.msn.com} \\
    \rev{www.dailymail.co.uk} \> \rev{www.pinterest.com} \\
    \rev{www.indeed.com} \> \rev{www.etsy.com} \\
    \rev{www.zillow.com} \> \rev{www.nypost.com} \\
    \rev{www.instructure.com} \> \rev{www.apple.com} \\
    \rev{www.discord.com} \> \rev{www.chaturbate.com} \\
    \rev{www.zoom.us} \> \rev{www.eporner.com} \\
    \rev{www.paypal.com} \> \rev{www.imgur.com} \\
    \rev{www.sharepoint.com} \> \rev{www.homedepot.com} \\
    \rev{www.bbc.com} \> \rev{www.ign.com} \\
    \rev{www.ups.com} \> \rev{www.craigslist.org} \\
    \rev{www.spotify.com} \> \rev{www.breitbart.com} \\
    \rev{www.fedex.com} \> \rev{www.roblox.com} \\
    \rev{www.theguardian.com} \> \rev{www.gamespot.com} \\
    \rev{www.weather.gov} \> \rev{www.chase.com} \\
    \rev{www.nextdoor.com} \> \rev{www.steamcommunity.com} \\
    \rev{www.usatoday.com} \> \rev{www.github.com} \\
    \rev{www.aliexpress.com} \> \rev{www.washingtonpost.com} \\
    \rev{www.temu.com} \> \rev{www.capitalone.com} \\
    \rev{www.samsung.com} \> \rev{www.bestbuy.com} \\
    \rev{www.patreon.com} \> \rev{www.wunderground.com} \\
    \rev{www.hulu.com} \> \rev{www.target.com} \\
    \rev{www.nih.gov} \> \rev{www.apnews.com} \\
    \rev{www.intuit.com} \> \rev{www.xfinity.com} \\
    \rev{www.canva.com} \> \rev{www.yelp.com} \\
    \rev{www.okta.com} \> \rev{www.healthline.com} \\
    \rev{www.adobe.com} \> \rev{www.lowes.com} \\
    \rev{www.character.ai} \> \rev{www.nbcnews.com} \\
    \rev{www.nfl.com} \> \rev{www.whatsapp.com} \\
    \rev{www.allrecipes.com} \> \rev{www.duosecurity.com} \\
    \rev{www.steampowered.com} \> \rev{www.slickdeals.net} \\
    \rev{www.realtor.com} \> \rev{www.costco.com} \\
\end{tabbing}

\end{document}